\documentclass[]{article}
\usepackage{emulateapj}
\usepackage{psfig}

\advance \voffset by 0.00cm\relax

\begin{document}

\title{A Comprehensive Study of Binary Compact Objects as 
       Gravitational Wave Sources: Evolutionary Channels, 
       Rates, and Physical Properties}

 \author{Krzysztof Belczynski\altaffilmark{1,2,3,4}, Vassiliki
         Kalogera\altaffilmark{1,2} and Tomasz Bulik\altaffilmark{3}}

 \affil{
     $^{1}$ Northwestern University, Dept. of Physics \& Astronomy,
        2145 Sheridan Rd., Evanston, IL 60208\\
     $^{2}$Harvard--Smithsonian Center for Astrophysics,
     60 Garden St., Cambridge, MA 02138;\\
     $^{3}$ Nicolaus Copernicus Astronomical Center,
     Bartycka 18, 00-716 Warszawa, Poland;\\
     $^{4}$ Lindheimer Postdoctoral Fellow\\
     belczynski, vicky@northwestern.edu, bulik@camk.edu.pl}

 \begin{abstract} 

A new generation of ground-based interferometric detectors for
gravitational waves is currently under construction or has entered the
commissioning phase (LIGO, VIRGO, GEO600, TAMA). The purpose of these
detectors is to observe gravitational waves from astrophysical sources
and help improve our understanding of the source origin and physical
properties. In this paper we study the most promising candidate sources
for these detectors: inspiraling double compact objects. We use
population synthesis methods to calculate the properties and coalescence
rates of compact object binaries: double neutron stars, black
hole-neutron star systems and double black holes. We also examine the
formation channels available to double compact object binaries. We
explicitly account for the evolution of low-mass helium stars and
investigate the possibility of common-envelope evolution involving
helium stars as well as two evolved stars. As a result we identify a
significant number of new formation channels for double neutron stars,
in particular, leading to populations with very distinct properties. We
discuss the theoretical and observational implications of such
populations, but we also note the need for hydrodynamical calculations
to settle the question of whether such common-envelope evolution is
possible. We also present and discuss the physical properties of compact
object binaries and identify a number of robust, qualitative features as
well as their origin.  Using the calculated coalescence rates we compare
our results to earlier studies and derive expected detection rates for
the Laser Interferometer Gravitational-wave Observatory (LIGO). We find
that our most optimistic estimate for the first LIGO detectors reach a
couple of events per year and our most pessimistic estimate for advanced
LIGO detectors exceed $\simeq 10$ events per year.

 \end{abstract}

\keywords{binaries: close --- gravitational waves --- stars: evolution,
formation, neutron, black hole}

\section{INTRODUCTION}

Gravitational waves are a natural consequence of Einstein's theory of
general relativity (Einstein 1916, 1918). Indirect evidence for their
existence came first from observations of the orbital decay of the
Hulse-Taylor binary pulsar (Hulse \& Taylor 1974, 1975a, 1975b; Taylor
\& Weisberg 1982, 1989). Direct detection though and analysis of
gravitational-wave sources are expected to provide a unique insight to
one of the least understood of the fundamental forces. They will also
allow us to investigate the physical properties of astronomical objects
that so far have been elusive because they do not emit any
electromagnetic radiation (e.g., double black holes in isolation).

A number of interferometers designed for gravitational-wave detection
are currently in operation, being developed, or planned.  First results
from the Japanese instrument TAMA300 have already been reported (Tagoshi
et al. 2001).  More advanced ground-based observatories are under
development or in the commissioning phase, i.e., the US Laser
Interferometer Gravitational-wave Observatory (LIGO; Abramovici et al.
1992), the British/German GEO-600 Observatory (Danzmann et al. 1995), the
French/Italian VIRGO project (Caron et al. 1995), and the Australian
ACIGA project (Sandeman 1998).  Also future space missions are being
planned, such as Laser Interferometer Space Antenna (LISA; Bender et al.
2000), a joint project of ESA and NASA with the launch planned for 2009.

Astrophysical sources of gravitational radiation relevant to
ground-based interferometers include:  inspiraling double compact
objects, binary stars, rotating neutron stars, neutron star
instabilities, supernovae, supermassive black holes and stochastic
background (for a review see Thorne 1987). Some of the most promising
candidates are the inspiral and coalescence of double compact objects 
(DCO), such as NS-NS, BH-NS, and BH-BH binaries. 
Successful detection of these sources at
reasonable event rates depends not only on the instrument sensitivity and
the strength of the gravitational-wave signals, but also on the
coalescence rates and the physical properties of the sources out to the
maximum distances of reach of a given instrument.

Coalescence rates of double compact objects have been investigated by a
number of different groups. For NS-NS binaries the rates can be
calculated in two ways: (1) theoretically, based on the predictions of
binary population synthesis calculations (e.g., Bethe \& Brown 1998;
Portegies Zwart \& Yungelson 1998;  Bloom, Sigurdsson \& Pols 1999;
Belczynski \& Bulik 1999; Fryer, Woosley \& Hartmann 1999) or (2)
empirically, based on the observed sample of Galactic binary pulsars
(e.g., Narayan, Piran, \& Shemi 1991; Phinney 1991;  Curran \& Lorimer
1995; Arzoumanian, Cordes \& Wasserman 1998; Kalogera et al.\ 2001; Kim, Kalogera, \&
Lorimer 2001). At present both methods appear burdened with
significant uncertainties, the theoretical approach due to the many
poorly constrained evolutionary model parameters, and the empirical
estimates due to small-number sample of observed NS-NS systems (for a
more detailed discussion see Kalogera et al.\ 2001). Predictions for
BH-NS and BH-BH binaries can be obtained only from theoretical
calculations, since no such systems have yet been observed. Accurate
predictions for BH binaries are even harder to obtain, as the evolution
of high-mass stars still challenge our understanding. Several results of
population synthesis calculations have been presented in the literature,
for BH-NS systems: Fryer et al.\ (1999); Belczynski, Bulik \&
Zbijewski (2000); Portegies Zwart \& Yungelson (1998), and for BH-NS and
BH-BH systems: Lipunov, Postnov \& Prokhorov (1997); De Donder \&
Vanbeveren (1998); Nelemans, Yungelson \& Portegies Zwart (2001b).

Given the history of population synthesis calculations, one may wonder
what is the purpose on another set of models of the formation of double
compact objects. The primary goal of this present study is to investigate
the relevant formation processes in the light of recent developments in
the evolution of compact object progenitors in binaries, and to clearly
identify classes of formation channels, their origin, and their relative
contributions. We focus not just on predicted rates, but at some level
more importantly on the physical properties of formed binaries. Our
parameter study is targeted to uncovering the systematic uncertainties as
well as identifying those (qualitative) properties that appear to be
robust. Investigation of these properties will help in characterizing the
range of possible gravitational-wave signals and identification of their
origin will allow reevaluating our results in the future as our
understanding of model unknowns improves. As part of our analysis we also
examine the implications of common-envelope phases involving helium stars
and as a result we identify a number of new formation channels with
important implications for the properties of double compact
objects\footnote{After the submission of this paper to {\em ApJ}, an early
paper by Tutukov \& Yungelson (1993) came to our attention.  In this paper
the authors allow for mass transfer from helium stars without a merger,
although not through CE evolution. There is qualitative agreement between
their and our results, however we point out that there are many
differences in the binary evolution modeling (e.g., Tutukov \& Yungelson
do not account for any kicks to NS or BH not for hypercritical accretion,
and many more). We also found out that CE evolution for low-mass helium
stars has been discussed qualitatively in the past, in the context of the
formation of NS-NS systems and X-ray binaries (van den Heuvel 1992; Taam
1996).}. Even though stellar structural characteristics may tentatively
support such common-envelope phases late in the evolution of helium stars,
we note that detailed hydrodynamical calculations will be needed to assess
the viability of such phases involving helium stars or pairs of evolved
stars as it has been assumed also in earlier studies (see Fryer et al.\
1999; Nelemans et al.\ 2001a)

Our paper is organized as follows:  In \S\,2, we present a detailed
description of our population synthesis code (named {\em StarTrack}).  
In \S\,3 we: describe the extent of our parameter study (\S\,3.1),
discuss the statistical accuracy of our calculations (\S\,3.2), present
the formation paths of double compact objects (\S\,3.3), and analyze in
detail their properties (\S\,3.4) and coalescence rates (\S\,3.5) for a
large number of different models.  In \S\,4, we discuss the implications
of the most important of our results and conclude with prospects for
gravitational-wave detection by ground-based interferometers.

\section{POPULATION SYNTHESIS MODEL}

\subsection{Single-Star Evolution}

\subsubsection{Overview}

In all the population synthesis models, we employ the analytic formulae
derived by Hurley, Pols \& Tout (2000, hereafter HPT)  to model the
evolution of single stars.  These formulae represent fits to results
from a large number of stellar evolution models calculated by Pols et
al.\ (1998), using the Eggleton stellar evolution code (Eggleton
1971,1972,1973;  latest updates and input physics described in Han,
Podsiadlowski \& Eggleton 1994 and Pols et al.\ 1995).

The HPT formulae\footnote{We implemented the formulae into a $C$ code,
compared the results of our implementation with the original HPT
$Fortran$ subroutines for ZAMS masses 0.5--$100\,{\rm M}_\odot$ and for
metallicities:  $Z=0.0001 - 0.03$, and found perfect agreement.} allow
us to calculate the evolution of a star given its Zero Age Main Sequence
(ZAMS) mass and its metallicity ($Z$).  Stellar evolution is followed
from ZAMS through a sequence of evolutionary phases depending on the
initial (ZAMS) stellar mass:  Main Sequence (MS), Hertzsprung Gap (HG),
Red Giant Branch (RG), Core Helium Burning (CHeB), Asymptotic Giant
Branch (AGB), which is further divided into Early AGB (EAGB) and
Thermally Pulsing AGB (TPAGB), and for stars stripped off their
hydrogen-rich layers: Helium Main Sequence (HMS), Helium Giant Branch
(HGB).  We end the evolutionary calculations at the formation of a
stellar remnant:  a white dwarf (WD), a neutron star (NS) or a black
hole (BH).  During the evolution we can track some of the basic stellar
parameters: radius, luminosity, stellar mass, and core mass.

These single star models include the effects of mass loss due to stellar
winds as described in HPT. Mass loss rates are adopted from the
literature for different evolutionary phases. For H-rich stars on MS
(Nieuwenhuijzen \& de Jager (1990); using $Z$ dependence of Kudritzki et
al.\ 1989), for RGB (Kudritzki \& Reimers 1978) and AGB (Vassiliadis \&
Wood 1993) stars, and for Luminous Blue Variables (HPT).  For He-rich
stars W-R, mass loss is included using rates derived by Hamann,
Koesterke \& Wessolowski (1995) and modified by HPT. Given the
importance of single star winds in the formation of compact objects
(e.g., Brown, Weingartner, \& Wijers 1996; Ergma \& van den Heuvel 1998;  
Fryer \& Kalogera 2001), we examine this element of single star
evolution and calculate synthesis models with varying strengths of stellar
winds.

We have introduced two modifications to the HPT formulae concerning the
treatment of (i) Helium-star evolution, and (ii) final remnant masses.
These are described in detail in the next two subsections.

\subsubsection{Helium Star Evolution}

After core helium exhaustion low-mass helium stars expand significantly
and develop a ``giant-like'' structure with a clearly defined core and a
convective envelope (Delgado, \& Thomas 1981;  Habets 1987; Avila-Reese
1993; Woosley, Langer, \& Weaver 1995). Radial expansion is very
important for stars in binaries as it may lead to mass transfer
episodes. Low-mass evolved helium stars with convective envelopes
possibly transfer mass on a dynamical time scale, and as a consequence a
common envelope (CE) phase may ensue\footnote{We note that whether a
common-envelope phase develops as well as the outcome depends not only
on the presence of convective envelopes but also on the detailed stellar
structure (density profile, masses of the envelope relative to the core)
as well as the mass ratio in the binary and the spin evolution of the
envelope during the mass-transfer episode (Fryer 2001; Rasio 2001; Taam
2001, private communications). The hydrodynamics of mass transfer from
helium stars with appropriate masses and companions have never been
studied so far, but such studies will be necessary to determine the
outcome of these evolutionary phases. In the present study we account
for such a possibility and examine its implication for the DCO 
populations.}. On the other hand, relatively massive helium stars do not
develop deep convective envelopes and do not experience any significant
radial expansion. Even if they happen to initiate mass transfer (MT), 
they respond differently to mass loss and a CE phase is not likely.

Based on the HPT formulae, the implied upper mass limit for helium stars
to develop deep convective envelopes is $M_{\rm conv} \sim
2.2\,M_\odot$), which we consider to be rather low. We have examined in
detail models of evolved helium stars (Woosley 1997, private
communication) and found that helium stars below 4.0\,M$_\odot$ have
deep convective envelopes, whereas slightly more massive helium stars
($\sim$ 4--4.5\,M$_\odot$) still form convective envelopes, although
shallower.  As an example, an evolved helium star of $M=2.5\,M_\odot$
develops a deep convective envelope reaching down almost to the core
through 90\% of the stellar radius.  Based on this, for our standard
model, we adopt $M_{\rm conv} = 4.5\,M_\odot$ for an evolved helium star
to develop a deep convective envelope. However, in face of the
uncertainties and the importance of helium-star evolution, we treat this
value as a model parameter and include it in our parameter study.
Throughout this paper we refer to the evolved helium stars with $M \leq
M_{\rm conv}$ as low-mass helium stars/giants (or low-mass HGB stars)
and to the stars with $M > M_{\rm conv}$ as massive helium stars/giants
(or massive HGB stars). We note that, apart from the issue of convective
envelopes, a comparison in terms of radial evolution alone between the
HPT formulae and other published models of helium stars (e.g. Habets
1987) shows excellent agreement, so we adopt these formulae in
calculating stellar radii. Detailed discussion of helium star evolution
in context of double compact objects formation is given in Belczynski \&
Kalogera (2001).

\subsubsection{Remnant Masses}

The masses of NS and BH calculated by HPT are {\em very} small.  For
progenitors of $M_{\rm zams}=40-100\,{\rm M}_\odot$, they obtain final
remnant masses in the range $\simeq 1.8-2.0\,{\rm M}_\odot$, (see their
Figure 20). On the other hand, measurements of BH masses in binaries,
although still highly uncertain, yield much higher values: $\sim$
3--$20\,{\rm M}_\odot$ (Orosz et al. 2001; McClintock et al. 2001; Froning
\& Robinson 2001; Wagner et al. 2001; Table 1 of Fryer \& Kalogera 2001
and references therein). The HPT analytic formulae are based on the
stellar models evolved at least to the formation of a carbon-oxygen (CO)
core, and these core masses are in good agreement with earlier stellar
models (e.g., Schaller et al. 1992).  Therefore, instead of using their
somewhat arbitrary prescription (see their equation [75]) for NS and BH
masses, we adopt a different formula derived on the basis of core-collapse
hydrodynamical calculations (Fryer 1999, Fryer \& Kalogera 2001). We note 
that there are still many uncertainties associated with the physics of the 
supernova mechanism, and there is still disagreement among research groups 
about the cause of the explosion (e.g., Liebendoerfer et al.\ 2001 and 
references therein). For our study we choose the specific core-collapse 
models, primarily because they are the only one available in the 
literature that allow us to calculate the remnant mass taking into account 
fallback, in a self-consistent way. 

\begin{figure*}[t]
\centerline{ \psfig{file=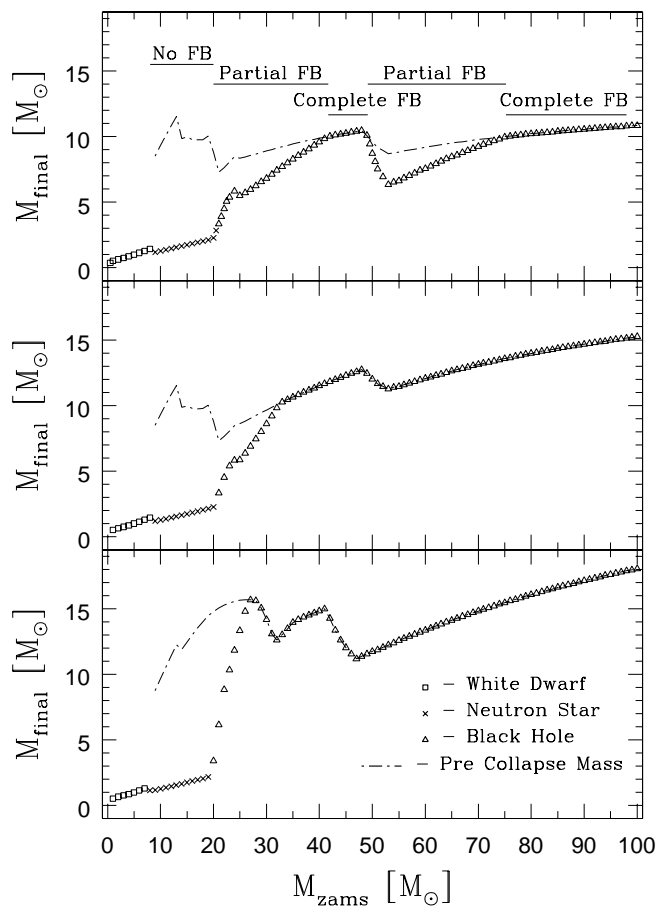,width=0.5\textwidth} }  
\caption{
Final remnant masses shown as a function of
progenitor ZAMS mass, for single Population I stars ($Z=0.02$) and for
three different wind mass-loss rates. Top panel: standard HPT wind
mass-loss rate. Middle panel: Wolf-Rayet wind rates decreased by two.
Bottom panel: all wind mass-loss rates decreased by a factor of two.
Different symbols represent different remnant types (square: white
dwarf; cross: neutron star; triangle: black hole). Stellar masses just
prior to the collapse are shown by the dot-dashed line. In the top
panel, we also mark the ranges of initial masses, for which NS or BH are
formed without any fall back, and with partial or complete fall back in
core-collapse events of massive stars. In all panels, a maximum NS mass
of $3\,{\rm M}_\odot$ has been assumed. Masses of WD remnants are
calculated as in HPT.}
\label{fig1}
\end{figure*}

We adopt the HPT formulae to calculate the final CO core mass of a star,
and we use stellar models of Woosley (1986) to obtain a final FeNi core
mass as a function of the CO core mass.  At the time of core collapse,
we describe the star with its current mass $M$, the final CO core mass
$M_{\rm CO}$ and the final FeNi core mass $M_{\rm FeNi}$. To estimate
the mass of the remnant formed, we follow the results of hydrodynamical
calculations of core collapse (Fryer 1999).  These calculations have
shown that (i) progenitors with $M_{\rm zams} \leq 20\,{\rm
M}_\odot$ do not experience any significant fall back and apart from the
collapsing core, the stellar outer layers are expelled as SN ejecta,
(ii) for progenitors with $M_{\rm zams} \geq 42\,{\rm M}_\odot$,
the whole pre-collapse star directly implodes to form a BH and no SN 
event
takes place, (iii) for intermediate progenitor masses remnants are
formed through an initial core collapse and subsequent partial fall
back. The outcome of a given core-collapse event is found to depend
primarily on the core mass of the collapsing star and not on the initial
(ZAMS) stellar mass. Taking into account that the stellar population
models used by Fryer (1999) were of constant mass (without wind or mass
transfer effects; see Fryer \& Kalogera 2001), we have developed a
prescription based on CO core masses. Using the HPT formulae we obtain
$M_{\rm CO} = 5\, {\rm M}_\odot$ for $M_{\rm zams} = 20\,{\rm M}_\odot$,
and $M_{\rm CO} = 7.6\, {\rm M}_\odot$ for $M_{\rm zams} \leq 42\,{\rm
M}_\odot$, for $Z=0.02$ using the standard wind mass loss prescription.
We then calculate the mass of the remnant as follows:  
 \begin{equation}
 M_{\rm rem}=\left\{ \begin{array}{lr} M_{\rm FeNi}& M_{\rm CO} \leq
5\,{\rm M}_\odot\\ M_{\rm FeNi} + f_{\rm fb} (M-M_{\rm FeNi})& 5<M_{\rm
CO}<7.6\\ M& M_{\rm CO} \geq 7.6\,{\rm M}_\odot\\ \end{array}\right.  
\label{Mrem}
 \end{equation} 
where $f_{\rm fb}$ is the {\em fall-back} factor, i.e.
the fraction (from 0 to 1) of the stellar envelope that falls back.
In Figure~\ref{fig1}, we present the initial-remnant mass relation, 
for single Population I stars ($Z=0.02$), and for three different 
wind mass-loss rates.

The top panel of Figure~\ref{fig1} shows the initial-remnant mass relation for
our standard prescription of wind mass-loss rates.
 
Stars of $M_{\rm zams}=$\, 0.5--0.8 ${\rm M}_\odot$ form helium white
dwarfs (He WD), of $M_{\rm zams}=$\, 0.8--6.3 ${\rm M}_\odot$ form
carbon-oxygen white dwarfs (CO WD), and of $M_{\rm zams}=$\, 6.3--8.0
${\rm M}_\odot$ form oxygen-neon white dwarfs (ONe WD).

For stars with initial mass $M_{\rm zams}$ in excess of 8 ${\rm
M}_\odot$, the stellar core mass exceeds the Chandrasekhar mass and once
nuclear reactions stop the core collapses into a NS. For $M_{\rm
zams}=8-20.0\, {\rm M}_\odot$, the outer stellar layers (on average
$\sim 7\, {\rm M}_\odot$) are lost in SN explosion, and no fall back
takes place.  This part of the initial-remnant mass relation is
characterized by a slow increase of the NS mass with increasing $M_{\rm
zams}$, as more massive progenitors form only slightly more massive
cores, which eventually collapse to NS.

Partial fall back is initiated above $M_{\rm zams}=20.0\, {\rm M}_\odot$
and the amount of material accreted back onto the collapsing core
increases linearly with increasing progenitor mass.  Due to the fall
back the initial-remnant mass relation significantly steepens over
$M_{\rm zams}=20.0\, {\rm M}_\odot$. Note that the remnant mass reaches
maximum NS mass $M_{\rm max}^{\rm NS}=3.0\,{\rm M}_\odot$ at $M_{\rm
zams}=20.7\, {\rm M}_\odot$, so in this model the most massive NS are
formed through fall back.

For $M_{\rm zams} \sim 25\, {\rm M}_\odot$ there is a small drop and a
subsequent flattening of the initial-remnant mass relation. At this and
higher progenitor masses stellar winds strip star completely of its
hydrogen layers. Once a naked helium star is formed the wind mass loss
rate increases significantly. The effect of wind mass loss enhancement
with increasing $M_{\rm zams}$ diminishes both the final core mass and
the pre-collapse envelope mass (the fall back mass reservoir) and
therefore decreases the remnant mass. With increasing $M_{\rm zams}$ the
fall back factor $f_{\rm fb}$ goes up, but because of the helium star
mass loss rate effect on the pre-collapse star, the remnant mass does
not increase as fast as before.

Stars more massive than $M_{\rm zams}=42\, {\rm M}_\odot$ collapse
directly into a BH at the end of their evolution. The initial-remnant
mass relation flattens out even more. This is a consequence of a
``fall-back saturation''. The fall back is now complete ($f_{\rm fb}=1$)
and cannot increase any more with progenitor mass, as it was the case
for less massive stars in the range of partial fall back. Now the
remnant mass depends only on the pre-collapse mass, which increases
rather slowly with progenitor mass as the wind mass loss rate increases
also with $M_{\rm zams}$.

Stars over $M_{\rm zams} \sim 48\, {\rm M}_\odot$ reach high
luminosities and radii exceeding the Humphreys-Davidson limit (Humphreys
\& Davidson 1994) and enter the Luminous Blue Variable (LBV) phase.
During the LBV phase stars lose mass at a very high rate, which result
in a sudden drop in the initial-remnant mass relation. Due to this
significant mass loss, stars over $M_{\rm zams}=49\, {\rm M}_\odot$ form
smaller cores and experience partial fall back at core collapse. The
effect of LBV-like mass loss takes over the effect of an overall
increase of CO core mass (and thus remnant mass) with $M_{\rm zams}$,
and it influences the remnant mass in two ways. First, LBV-like
mass-loss rates increase with increasing stellar mass leading to a
decrease of remnant masses. Second, for higher stellar ZAMS masses, the
LBV phase begins earlier in the life of a star. Hence, the stars 
are stripped of their
hydrogen layers earlier, and less massive helium stars are formed,
leading to a decrease in remnant masses.  In the standard wind model,
remnant masses reach their local minimum at $M_{\rm zams}=52.5\, {\rm
M}_\odot$, when the LBV phase sets in the earliest possible, i.e., at
the beginning of Hertzsprung gap. For more massive stars, the overall
increase of remnant mass with progenitor mass dominates once again, as
CO core masses steadily grow and the contribution of fall back steadily
increases.

For progenitors with $M_{\rm zams} \geq 72.3\, {\rm M}_\odot$ remnants
are once again formed through complete fall back and the initial-remnant
mass relation flattens out, following the slow increase of pre-collapse 
mass with $M_{\rm zams}$.

In the middle panel of Figure~\ref{fig1}, results are shown for Wolf-Rayet
mass-loss rates reduced by a factor of two. The initial-remnant mass
relation changes only slightly from that of the top panel as 
partial fall back at high ZAMS masses is eliminated. As expected,
compact objects are slightly more massive and the maximum BH mass
increases from $\sim 11 \, {\rm M}_\odot$ (top panel) to $\sim 15 \,
{\rm M}_\odot$ (see also Wellstein \& Langer 1999).

In the bottom panel of Figure~\ref{fig1}, results are shown for the case of all
wind mass-loss rates reduced by a factor of two, and we find some
significant changes in the initial-remnant mass relation.  First, as
expected, the compact object masses are higher than for both models
described above; the maximum mass of the compact object in this model
increases to $\sim 19 \, {\rm M}_\odot$. Second, the qualitative shape
of the relation changes. As in the standard wind model, we find two
sudden drops in the relation, with the first one being more prominent
than before. In the standard model the first drop corresponds to the
stripping star of its hydrogen layers and the formation of a helium
star, whereas the second drop is related to the onset of LBV phase for
very luminous and extended stars. These two effects lead to
significantly increased mass loss rates, and thus decrease the remnant
masses. Detailed examination of this model reveals that in the bottom
panel, the first drop corresponds to the onset of the LBV phase and the
second to the helium star formation.  Since all wind rates are reduced
and wind mass loss rates depend strongly on the stellar mass, helium
stars are exposed only for stars of higher initial mass. Consequently,
stars with much lower initial masses (below $M_{\rm zams} \sim 40\, {\rm
M}_\odot$) retain their hydrogen-rich layers and are able to evolve to
higher luminosities and larger radii, which shifts the onset of the LBV
phase to lower initial masses. Another difference is that, the immediate
progenitors of compact objects tend to be more massive, and therefore
direct BH formation spans much wider range of initial masses than for
the standard wind model. These more massive progenitors form more
massive CO cores, and none of the factors that decrease pre-collapse
masses are strong enough to decrease the CO core masses below $7.6\,
{\rm M}_\odot$ (see eq.~\ref{Mrem}). Consequently, BH formation through partial
fall back does not occur for two separate $M_{\rm zams}$ ranges (as is
the case for $M_{\rm zams}$= 21--42 and 48--72$\, {\rm M}_\odot$ in the
standard wind model).

\subsection{Binary Evolution}

\subsubsection{Overview}

We use Monte Carlo techniques to model the evolutionary history of double
compact objects and study their physical properties. We generate a large
number ($N \geq 10^6$) of primordial binaries, drawing their initial
physical parameters (component masses, orbital separations and
eccentricities) from assumed distributions described in \S\,2.2.2. Besides
that, Monte Carlo events are used to {\em i)} set the time of birth of
each binary in the disk of our Galaxy (for constant star formation rate
the probability distribution of the birth time is flat in the range
$0-10$\,Gyr), {\em ii)} set the kick velocity direction and magnitude each
NS or BH receives when formed in an asymmetric SN explosion and {\em iii)}
set the position along the binary orbit in which SN take place. In all
other instances, the evolution is modeled based on a set of formulae and
prescriptions (described in what follows) that depend on the binary 
properties.

The evolution of each system is followed in a number of time steps. The
time step is chosen to be a small fraction ($\leq 0.01$) of the
evolutionary phase lifetime of the more rapidly evolving stellar
component, making sure that the relative change in radius is always less
than $10$\%.  We start the evolution of a given binary with two components
at ZAMS, and we stop the calculation when the two components have formed
stellar remnants or when a merger occurs or when the system has reached
the present time, i.e., the sum of its age and its birth time is equal to
10\,Gyr).  The evolution of each binary component in isolation is
calculated as described in \S\,2.1, and a number of evolution effects on
the binary orbit (e.g., mass and angular momentum losses due to stellar
winds) are taken into account.  At every evolutionary time step, we check
for possible binary interactions. If any of the components fills its Roche
lobe, we calculate the effects of mass transfer and possible mass and
angular-momentum loss on the orbit.  Depending on the masses of the
components and their evolutionary stage, we apply appropriate
prescriptions for the modeling of different mass transfer events (as
described in \S\,2.2.4).  If the binary survives the mass-transfer event
(i.e., both stellar components fit within their Roche lobes), we continue
to model its evolution. At every time step, we also check the single-star
models (see \S\,2.1) to determine if any of the components has finished
its nuclear evolution, and has turned into a compact remnant. If a remnant
is born in a SN explosion, we calculate the effects of SN kicks and mass
loss on the binary orbit (both circular and eccentric). If a binary is
disrupted in SN explosion, we follow the single-star evolution of
components, otherwise we continue with the evolution of the whole binary.
Once a binary consists of two remnants, we calculate its merger lifetime,
i.e., the time until the components merge due to gravitational radiation
and associated orbital decay, and we study the properties and formation
rates of different classes of binaries containing compact objects.  In
what follows we describe in detail prescriptions for binary evolution and
all model assumptions as chosen for our standard model.  The assumptions
for models in our parameter study are given in \S\,3.1.

\subsubsection{Distributions of Initial Parameters}

A binary system is described initially by four parameters: the mass of
the primary $M_1$ (the initially more massive component), the mass ratio
$q={M_2 \over M_1}$, where $M_2$ is the mass of the secondary (initially
less massive component), the semi-major axis $A$ of the orbit, and the
orbital eccentricity $e$. We assume that the initial distributions of
these parameters are independent.

For both, single stars and binary system primaries, we adopt the initial
mass function derived by Scalo (1986),
 \begin{equation} 
 \Psi(M) \propto {M}^{-2.7},
 \label{imf}
 \end{equation}
 in the mass range $M_1=5-100\, {\rm M}_\odot$ relevant
to compact object formation. Although, for single stars, the minimum
initial mass for NS formation is $\simeq 8\, {\rm M}_\odot$ (or maybe
higher), a lower limit of $5\, {\rm M}_\odot$ ensures that we do not
miss any NS progenitors due to binary evolution. Mass transfer in a
binary can increase any component's mass, and effectively decrease the
minimum initial mass of NS progenitors. Let us consider a border-line
case of the lowest-mass primary ($5\, {\rm M}_\odot$) and the
corresponding highest-mass secondary ($5\, {\rm M}_\odot$). The primary
evolves off the MS with a helium core of $\sim 1\, {\rm M}_\odot$ and
$\sim 4\, {\rm M}_\odot$ in its envelope. For our standard model of
binary evolution, we assume that only half of the donor
envelope can be accreted by its companion (see \S\,2.2.4). 
So for this border-line case,
the initial secondary mass can be increased to $\leq 8\, {\rm
M}_\odot$, or for all other cases, to masses smaller than the mass limit
for NS formation.

Following Kuiper (1935), we assume a flat mass ratio distribution,
 \begin{equation}
 \Phi(q) = 1 
 \end{equation} 
 in the range $q=0-1$.  Given value of the primary mass
and the mass ratio, we obtain the mass of the secondary $M_2= q M_1$. We  
evolve only systems with secondaries of $M_{\rm zams} \geq 0.5\, {\rm
M}_\odot$, as less massive stars will not evolve within a Hubble time.
 
The distribution of the initial binary separations is assumed to be 
flat in the logarithm (Abt 1983),
 \begin{equation}
 \Gamma(A) \propto {1\over A},
 \end{equation} 
 where $A$ ranges from a minimum value, such that the
primary fills its Roche lobe, up to $10^5 \, {\rm R}_\odot$.

We adopt the thermal-equilibrium eccentricity distribution for initial
binaries,
 \begin{equation}
 \Xi(e) = 2e,  
 \end{equation} 
 in the range $e = 0-1$ (e.g., Heggie 1975; Duquennoy \& Mayor 1991).

\subsubsection{Evolution of Binary Orbit}

Tidal circularization of binary orbits takes place in systems in where
the size of any component is comparable to the binary separation (Zahn
1978).  This condition is satisfied when at least one of the binary
components evolves beyond the MS or the initial binary separation is
small.  To calculate the effects of tidal circularization on a binary
orbit we follow the prescription developed by Portegies Zwart \& Verbunt
(1996):  circularization occurs, if the stellar radius of one component
is larger than $0.2$ of the periastron binary separation.  The orbital
elements ($A,e$) change under conservation of angular momentum until the
new periastron distance is equal to 5 stellar radii of the component
driving the circularization, or until the orbit is circularized ($e=0$).  
We also assume that tidal circularization takes place instantaneously
(i.e. in one evolutionary time step).

Stellar winds affect binary orbits through mass and angular-momentum
losses. Assuming a spherically symmetric stellar wind, which carries
away the specific angular momentum of the mass-losing star, and a
circular orbit (Jeans-mode mass loss), the change in the binary
separation is given by
 \begin{equation}
 A (M_1+M_2) = {\rm const}. 
 \end{equation}
 For eccentric orbits the change in $A$ is rather similar (Vanbeveren, 
Van Rensbergen \& De Loore 1998).

Orbits are also affected by mass transfer events and SN explosions. The
treatment of these effects is described in the subsections that follow.

\subsubsection{Mass Transfer Events}

When one of the binary components fills its Roche lobe, mass is
transferred to its companion through the inner Lagrangian point. The
responses of both components and the orbit as well as the outcomes of MT
phases depend on the masses and evolutionary stages of the two stars. In
our calculations we distinguish between (i) dynamically stable, in
general non-conservative (allows for mass and angular-momentum loss from
the system) mass transfer, and (ii) dynamically unstable mass transfer
that leads to common envelope evolution (for more details and reviews
see, e.g., Ostriker 1975, Paczynski 1976, Iben \& Livio 1993, Rasio \&
Livio 1996;  Taam \& Sandquist 2000). We are interested in the end
products of MT events, and therefore, we assume that MT takes place
instantaneously. Following any MT event, we update the binary component
masses and evolutionary stages and calculate their radii \footnote{For
bare CO cores, in the absence of any published models, we adopt a
fixed radius of 0.01 $R_\odot$. We find that variation of this assumed
value does not affect our results in any appreciable way (even for an 
increase of this radius by an order of magnitude).}.
Using the post-MT radii and separation, we examine whether systems
survived the MT event and a component merger was avoided.

{\em Non-conservative Stable Mass Transfer.} This phase is implemented
if (i) the donor is a MS or
HMS star transferring mass to an accretor of any 
evolutionary stage,
or (ii) the donor is a H-rich or He-rich giant and the accretor is not, 
and the following is true
 \begin{equation} 
 M_{\rm don} \leq c_{\rm r} M_{\rm acc}
 \label{cr}
 \end{equation} 
 where $M_{\rm don},M_{\rm acc}$ are the donor and accretor masses at
the beginning of the MT episode. Based on earlier results obtained by
Hjellming \& Webbink (1987), Kalogera \& Webbink (1996) and Ritter (1999), 
$c_{\rm r}=2.5$ if the donor is in HG and the accretor is a MS star, 
and $c_{\rm r}=1$ in all other cases.

During this type of MT episodes, part of the mass lost by donor ($f_{\rm
a}$) is accreted onto the companion, and the rest is lost from the
system with a specific angular momentum equal to ${2 \pi j A^2 / P}$
(Podsiadlowski, Joss \& Hsu 1992). 
The corresponding orbital separation $A$ change can be calculated 
for $f_{\rm a}=0$ from:

\begin{equation}
 {A_f \over A_i} = 
 {M_{\rm don}^f+M_{\rm acc}^f \over M_{\rm don}^i+M_{\rm acc}^i} 
 \left({M_{\rm don}^f \over M_{\rm don}^i}\right)^{2(j-1)} 
 \exp \left[ {2j(M_{\rm don}^f-M_{\rm don}^i) \over M_{\rm acc}^i} \right], 
\label{pod1}
\end{equation}
and for $f_{\rm a}>0$ from:
\begin{equation}
 {A_f \over A_i} = 
 {M_{\rm don}^f+M_{\rm acc}^f \over M_{\rm don}^i+M_{\rm acc}^i} 
 \left({M_{\rm don}^f \over M_{\rm don}^i}\right)^{c_1} 
 \left({M_{\rm acc}^f \over M_{\rm acc}^i}\right)^{c_2},               
\label{pod2}
\end{equation}
where
 $$c_1 \equiv 2 j (1-f_{\rm a})-2,$$
 $$c_2 \equiv - {2j \over f_{\rm a}} (1-f_{\rm a})-2,$$
 $$M_{\rm acc}^f=M_{\rm acc}^i+f_{\rm a}(M_{\rm don}^i-M_{\rm don}^f),$$
 and where $M_{\rm acc},M_{\rm don}$ indicate binary component -- donor
and accretor masses, and the indices $i,f$ denote the initial and final
values, respectively.  For our standard model, we assume that half of
the mass lost from the donor is also lost from the system ($1-f_{\rm
a}=0.5$) with specific angular momentum $j=1$.

For MS and HMS donors, we follow the MT episode in small ($<1$\%)  mass
increments and we update the stellar masses and radii as well as the
orbital separation and the Roche-lobe radii. Mass transfer is terminated
when, for both stars, the Roche-lobe radii exceed the stellar radii. We
also take into account possible rejuvenation of the accretor (as described 
in HPT) and update its radius and core mass. If during this mass
transfer phase the donor mass decreases below $0.5 \, {\rm M}_\odot$ for
MS stars or below $0.3 \, {\rm M}_\odot$ for HMS star, we terminate our
calculations for the specific system, since such low-mass stars will not
end their evolution in a Hubble time.  For donors beyond the MS, we
assume that the entire stellar envelope is lost and the final donor mass
is equal to its core mass: He core or CO core mass for H-rich or He-rich
stars, respectively.

The mass-ratio criterion for stable mass transfer (see eq.~\ref{cr}) also
accounts for cases where the accretor is a NS or BH (see Kalogera \&
Webbink 1996; Kalogera 2000). We extend this treatment also to WD and
allow for non-conservative mass transfer in cases where the criterion is
satisfied. When WD masses exceed 1.44 ${\rm M}_\odot$ (Chandrasekhar mass
according to our stellar models) we assume that a Type Ia SN
occurs\footnote{The issue of progenitors of Type Ia SN is still not
resolved (e.g., Branch et al.\ 1995). However, our assumption of the
occurrence of Type Ia SN does not influence our results. Since accreting
WD companions are less massive or of comparable mass, even if WD were to
collapse to NS due to accretion, the chance of forming double compact
objects is vanishing} We also allow for accretion-induced collapse of NS
to BH and we treat the boundary $M_{\rm max}^{\rm NS}$ as a model
parameter. It is clear that our treatment of dynamically stable accretion
onto compact objects does not include some important aspects (e.g.,
Eddington-limited accretion, transient systems), but it turns out that
these MT phases are not crucial to double compact object formation.

{\em Conservative Mass Transfer.} In general, we assume that dynamically
stable mass transfer is non-conservative and we include the case of
conservative mass transfer (no mass or angular-momentum loss from the
system) in our parameter study. In this case, $f_{\rm a}=1$ and 
equation~\ref{pod2} reduces to (e.g. Verbunt \& Van den Heuvel 1995),
 \begin{equation}
 {A_f \over A_i} = \left({ M_{\rm don}^i M_{\rm acc}^i \over M_{\rm 
don}^f M_{\rm acc}^f }\right)^{2}. 
 \end{equation}

{\em Standard Common Envelope Phase.} In all cases where criterion 2.7
is not satisfied, the donor has reached the giant branch, and the
accretor is either a MS, HMS, WD or a high mass He-rich giant branch star, 
we expect mass transfer to be dynamically unstable and a common envelope
spiral in and ejection to occur (Iben \& Livio 1993). We assume that,
during the CE phase, the donor loses its entire envelope and the
spiraling-in companion does not accrete any of the envelope material.

We base our treatment on the energy formalism of the CE evolution
(Webbink 1984), where the envelope is ejected on the expense of the
binary orbital energy, and a tight post-CE system forms.  The final
orbital separation $A_f$ is calculated using,
 \begin{equation}
 \alpha_{\rm ce} \left( {G M_{\rm don}^f M_{\rm acc} \over 2 A_{\rm f}} 
-
{G M_{\rm don}^i M_{\rm acc} \over 2 A_{\rm i}} \right) =
{G M_{\rm don}^i M_{\rm don,env} \over \lambda R_{\rm don,rl}}
 \label{ce}
 \end{equation}
 where, $M_{\rm don}$ and $M_{\rm acc}$ are masses of the donor and its
companion, $M_{\rm don,env}$ is mass of donor's envelope, $R_{\rm don,rl}$
is the Roche lobe radius of the donor, and the indices $i,f$ denote the
initial and final values, respectively. Parameter $\lambda$ describes the 
central concentration of the giant (de Kool 1990; Dewi \& Tauris 2000).
The right hand side of equation~\ref{ce} expresses the binding energy of
the donor's envelope, the left hand side represents the difference
between the final and initial orbital energy, and $\alpha_{\rm ce}$ is
the CE efficiency with which orbital energy is used to unbind the
stellar envelope. If the calculated final binary orbit is too small to
accommodate the two stars then a merger occurs. In our calculations, we 
combine $\alpha_{\rm ce}$ and $\lambda$ into one CE parameter, and for
our standard model, we assume that $\alpha_{\rm ce}\times\lambda = 1.0$.

{\em Double Common Envelope Phase.} Brown (1995) suggested that if two
binary components are giants with convective envelopes and mass transfer
is initiated, then a double CE can form, where the two stellar cores
spiral in the combined stellar envelopes. As in the case of low-mass
helium stars, hydrodynamical studies of such a phase have not been
undertaken, and we note that, despite its plausibility, the validity of
such an assumption has not been demonstrated in detail (see also footnote
in \S\,2.1.2). We use an analogue of equation~\ref{ce} to describe the
energy balance of a double CE and to calculate the change of binary
separation $A$,
 \begin{equation}
 \alpha_{\rm ce} \left( {G M_{\rm 1}^f M_{\rm 2}^f \over 2 A_{\rm f}} -
{G M_{\rm 1}^i M_{\rm 2}^i \over 2 A_{\rm i}} \right) = 
{G M_{\rm 1}^i M_{\rm 1,env} \over \lambda R_{\rm 1,rl}} +
{G M_{\rm 2}^i M_{\rm 2,env} \over \lambda R_{\rm 2,rl}}
 \end{equation}
 where $M_{\rm 1},M_{\rm 2}$ are the two donor masses, $M_{\rm 1,env},
M_{\rm 2,env}$ are their envelope masses and other symbols have the same
meaning as in equation~\ref{ce} (see also Nelemans et al.\ 2001a).

{\em Common Envelope with Hyper-critical Accretion.} It has been
suggested that if a NS or BH evolves through a CE phase, the compact
object may accrete significant amounts of material because of
hyper-critical accretion (Blondin 1986; Chevalier 1989, 1993; Brown
1995). Bethe \& Brown (1998) derived an analytic scheme for the
evolution of the binary orbit and the increasing mass of the accreting
compact object, under the assumption that the mass of the compact object
is much smaller than the mass of the giant.  However, for a large
fraction of CE events this assumption is not justified, e.g., binaries
with NS and a low-mass He-rich giant, or with a $\sim 5-10 \, {\rm
M}_\odot$ BH and a relatively massive giant companion.

In {\em StarTrack}, we relax the simplifying assumption made by Bethe \&
Brown (1998) and derive a formulation that can be applied to CE phases
with hyper-critical accretion for any compact object and companion
masses. This new formulation involves the numerical solution of a set of
ordinary differential equations for the final compact object mass and
orbital separation of the post-CE system.  This non-approximate
solution, results in tighter post-CE systems and lower mass compact
objects compared to the analytic expression of Bethe \& Brown (1998). 
We present the detailed derivation in Appendix A.

{\em Unmodeled cases.} There are two cases of binary interactions that
are not covered by the above MT prescriptions, as we cannot readily
foretell the outcome of some mass transfer events. We do not follow the
subsequent evolution of binaries going through such events, but we keep
a record of their occurrences.  For both types of unmodeled MT events,
the fraction of primordial binaries that encounter such events is below
0.1\% for all our models.

These two types of unmodeled events are:  (i) Unevolved donor with an
evolved companion (i.e., giant with convective envelope). This
configuration is encountered very rarely, as usually it is the giants
that fill their Roche lobes. (ii) Compact objects accreting from massive
evolved helium stars (with radiative envelopes). We expect that most of
these cases lead to stellar mergers.  Moreover, if any two of MS, HMS or
massive HGB stars fill their Roche lobes at the same time, the formation
of a contact system is recorded, and the calculations for these systems
are terminated\footnote{We find that the fraction of contact systems
formed is rather independent of the adopted evolutionary model and is in
the range 4--6\% of the total simulated binary population.}.

\subsubsection{Core Collapse and Supernova Explosion Events }

For each massive star, the time of core collapse is set by the
single star models (taking though into account mass variations due to
winds and binary interactions).  When either component of a binary
reaches this stage, we generate a random location in the orbit for the
event to take place (note that for eccentric binaries this choice will
affect the outcome, since the components' separation and relative
velocities are different at different locations in the orbit).  The 
core-collapse event is assumed to be instantaneous and the mass of the
remnant is calculated using equation~\ref{Mrem}. Note that if the remnant is
formed through complete fall back (leading always to direct BH
formation), we do not expect a SN explosion (hence no kick and no mass 
loss) and the
orbit remains unchanged (Fryer 1999). When BH is formed through partial 
fallback we treat the event as a SN explosion (see Podsiadlowski et al.\ 
2001). 

We calculate the effect of a SN event on binaries in three steps: (i)
We estimate the mass of the remnant. The rest of exploding star is
immediately lost from the binary (with the angular momentum specific to the
exploding component). We assume that the ejecta do not have
any effect on the companion (e.g., Kalogera 1996). (ii) We calculate the
compact object velocity which is the vector sum of the orbital velocity
of the pre-collapse star at the orbital position and the kick velocity.
The kick velocity is assumed to be randomly oriented and its magnitude
is drawn from an assumed distribution. The kick magnitude is
also scaled with the amount of material ejected in the SN explosion,
 \begin{equation} 
 V_{\rm kick}=(1-f_{\rm fb})*V, 
 \end{equation} 
 where $V$ is the kick magnitude drawn from a given assumed
distribution, and $f_{\rm fb}$ is a fall back parameter defined in
\S\,2.1.3, and the $V_{\rm kick}$ is the kick magnitude we use in our
calculations. For NS remnants and no fall back ($f_{\rm fb}=0$) and
$V_{\rm kick}=V$.  In our standard model we use a kick magnitude
distribution very similar to the one derived by Cordes \& Chernoff 
(1998): a weighted sum of two Maxwellians, one with $\sigma=175$\
km~s$^{-1}$ (80\%) and the second with $\sigma=700$~km~s$^{-1}$ (20\%).
This distribution accounts for the fact, that besides average velocity
pulsars a significant fraction of neutron stars have velocities above 
500 km\,s$^{-1}$ (Arzoumanian, Cordes, \& Chernoff 1997).
(iii) We calculate the total energy (potential and kinetic) of the new
orbit for the remnant (new velocity and mass, same relative position)
and its companion. If the total energy is positive, then the system is
disrupted, and its components will evolve separately. We calculate their
subsequent evolution as single stars, but we do not follow
their trajectories in the present study. If the total binary energy is
negative, the system after SN explosion is bound, and we calculate its
new parameters ($e$ and $A$). We also check whether the two components have
merged due to the SN mass loss and kick in which case we terminate the
evolution. Finally, we calculate the post-SN center of mass velocity of a
binary.

\subsubsection{Orbital Decay Due to Gravitational Radiation}

Once both components of a binary have ended their nuclear evolution and
formed stellar remnants, their orbits change only due to angular
momentum loss through gravitational radiation. We calculate the merger
times of these binaries using the formalism developed by Peters (1964),
the result of which is briefly described below.

The decay rates of eccentricity $e$ and orbital separation $A$ are
combined to derive the eccentricity evolution with time $e(t)$ and 
integrate to the limit $e\rightarrow 0$ (as the system approaches 
merging, i.e., $A\rightarrow 0$): 
 \begin{equation}
 t_{\rm merg} = {12 \over 19} {c_0^4 \over \beta} \times 
\int\limits_{0}^{e_0} { e^{29/19} [1+(121/304) e^2]^{1181/2299}\,de 
\over (1-e^2)^{3/2} },\\
\label{mergtim} 
\end{equation}
 where $e_0$ and $A_0$ are the initial eccentricity and orbital
separation, $$c_0 = A_0 {1-{e_0}^2 \over {e_0}^{12/19} [1+(121/304)
{e_0}^2]^{870/2299}},$$ $$\beta = {64 \over 5} {G^3 M_1 M_2 (M_1+M_2)
\over c^5},$$ and where $G$ is the gravitation constant and $c$ is the
speed of light. In practice the limit $A\rightarrow 0$ (accompanied by
$e\rightarrow 0$) is unrealistic, since a merger occurs earlier, but the
very final parts of this integration do not contribute significantly to
the merger time, as the very late stages of inspiral proceed on very
short timescales. For example, the merger time we calculate for 
PSR~B1913+16 is within 1\% of the result obtained by solving 
the system of two ordinary equations for the eccentricity and orbital
separation (e.g., Junker \& Schaefer 1992). 

We note that, for a circular orbit, we obtain a merger time:
 \begin{equation}
 t_{\rm merg} = {a_0^4/(4 \beta)}.
 \end{equation}

\subsubsection{Star Formation History and Binary Fraction}

For most of our models, we assume that star formation has been
continuous in the disk of our Galaxy for the last 10\,Gyr as inferred
from observations (e.g., Gilmore 2001) and as predicted by recent
theoretical modeling (e.g., Kauffmann, Charlot \& Balogh 2001).  We  
start the evolution of a single or a binary system $t_{\rm birth}$\ ago,
and follow it to the present. The birth time $t_{\rm birth}$ is drawn
randomly within the range 0--10\,Gyr. In our parameter study, we examine
a model of instantaneous star formation.  

For most of our models, we also assume a binary fraction of $f_{\rm
bi}=0.5$, which means that for any 150 stars we evolve, we have 50
binary systems and 50 single stars. We treat $f_{\rm bi}$ as a model
parameter and include it in our parameter study.  Finally, in all our 
models we assume solar metallicity $Z=0.02$.

\section{RESULTS}

\subsection{Population Synthesis Parameter Study}

We perform an extensive parameter study in order to assess the
robustness of our population synthesis results.  In Table~\ref{models}
we summarize all our models based on those assumptions that are
different from our standard model (model A). Parameter values and
adopted distributions in the standard model are as given in the two
previous sections: \S\,2.1 and \S\,2.2.

In models marked with letter B, we use different distributions of SN
kick magnitudes imparted to NS at birth.  In model B1 we assume
symmetric SN explosions, whereas in models B2--12 we draw kick
velocities $V_k$ from a single Maxwellian:
 \begin{equation}
 g(V_k) \propto {V_k}^2 \exp\left[{-({V_k} / \sigma)^2}\right],
 \end{equation} 
 varying $\sigma$ values in the range $10-600 {\rm km
s}^{-1}$. In model B13 we use a kick distribution of the form suggested
by Paczynski (1990):  
 \begin{equation} 
 f(V_k) \propto [1+(V_k/\sigma)^2]^{-1}, 
 \end{equation} 
 which allows for a significant fraction of low-magnitude kicks. We use
$\sigma = 600 {\rm km\,s}^{-1}$, which gives a reasonable fit to the
population of single pulsars in the solar vicinity (Hartman 1997).

In model C, compact objects (NS or BH) are not allowed to accrete any
material in CE events. In models D1--2, we reduce our conservative maximum
limit on the NS mass of 3 $M_\odot$ down to 2 and 1.5 $M_\odot$. Models
E1--3 present evolution with different effective CE efficiencies
($\alpha_{\rm ce}\times\lambda$).  In model F1, we significantly decrease
the amount of material accreted by companions in non-conservative MT
events, whereas in model F2, we examine the case of conservative mass
transfer.  In models G1--2, we vary the wind mass loss rates. In model G1,
we decrease the wind mass loss rate by a factor of 2, for all stars and at
all evolutionary stages, whereas in model G2, we enhance it by factor 2.
In other words, we calculate the wind mass loss rate as described in
\S\,2.1.1, and multiply it by $f_{\rm wind}$ given in Table~\ref{models}.  
Model H1 corresponds to evolution with $M_{\rm conv}=4.0 M_\odot$, the
upper mass limit for helium giants with convective envelopes that would
initiate a CE phase. Model H2 does not allow for CE evolution for any
Roche-lobe filling helium stars regardless of their mass (effectively
$M_{\rm conv}=0$). In model I, the star formation rate (SFR) is altered,
and instead of continuous we assume a burst-like star formation history,
with all stars being formed 10 Gyr ago.  In model J, we use an initial
mass function (IMF) with a Salpeter (1955)  exponent of $-2.35$, and in
models K1--2, we vary the binary fraction of binary systems in the
primordial stellar populations. In models L1 and L2, we vary the specific
angular momentum of material lost during dynamically-stable MT events. In
models M1 and M2, we use different distributions of initial binary mass
ratios. Model N corresponds to the artificial case of evolution without
allowing for the expansion of evolved helium stars.  This model is
nonphysical since low mass helium stars are known to evolve to giant-like
stages and to significantly increase their radii (for a detailed
discussion see Belczynski \& Kalogera 2001 and references therein). We
include this model in our parameter study only for the purposes of
comparing our results to earlier studies that did not not account for
helium star radial evolution. In model O, we increase the mass range of
compact object formation through partial fall back, resulting to no BH  
forming through a direct collapse.

\subsection{Statistical Accuracy}

Apart from the systematic uncertainties of our results which we examine
with the broad parameter study discussed above, we also examine the
statistical accuracy of our population synthesis models, to make sure
that we do not burden the quantitative predictions with unnecessary
uncertainties.  To determine the intrinsic statistical accuracy of our
synthesis models, we performed 30 different realizations of our standard
model (A), each with $N=10^6$ independently generated primordial
binaries. We decide to limit ourselves to $N=10^6$ runs, given our
computational resources.

In Table~\ref{stat}, we present the results we obtained by comparing the
30 different runs. For each of the three DCO populations, we list the
mean number of systems formed from the set of $N=10^6$ runs, its
standard deviation (as a percentage), as well as the maximum variation
in the predicted numbers of systems. We find that the standard
deviations do not exceed 10\% for any type of DCO, whereas the maximum
variation remains below 10\% for the most frequent NS-NS binaries and
below 30\% for the least frequent population of BH-NS binaries. In what
follows, we present results for our standard model using the combined
total of these 30 runs (or their subgroups) and therefore statistical
inaccuracies are greatly reduced and become insignificant compared to
the systematic uncertainties. For some of the other models in our
parameter study (e.g., E1, G2) we also considered results from runs with
more than $10^6$ primordial binaries to ensure that our statistical
errors remain well below 10\% for all three DCO types.

\subsection{Formation Paths of Double Compact Objects}

We consider double compact objects with NS or BH (NS-NS, BH-NS or
BH-BH binaries) with merger times shorter than $10^{10}$\ yr, thus {\em
coalescing} DCO.  In our standard evolutionary model, the population of
coalescing DCO is dominated by NS-NS systems (61\%), with a significant
contribution by BH-BH binaries (30\%), and a small contribution by BH-NS
objects (9\%). In what follows we discuss the main qualitative
characteristics of the important formation paths (with relative
formation frequencies higher than 1\%) as well as their origin.

In Table~\ref{channels}, we present the most important formation
channels of coalescing DCO, for our standard model.  Formation channels
of NS-NS, BH-NS and BH-BH binaries are marked by NSNS, BHNS and BHBH,
respectively, they are listed in order of decreasing relative formation
frequency (second column) with respect to the whole DCO coalescing
population.  The details of each evolutionary sequence, i.e., MT
episodes and SN explosions are also given.  Results were obtained based
on the evolution of $3\times 10^7$ primordial binaries.

\subsubsection{Populations of Double Neutron Stars}

Using the {\em StarTrack} population synthesis code, we identified a
number of new NS-NS formation channels (see below).  This is a result of
two improvements in the implementation of our population synthesis code,
since our previous short study of the NS-NS formation (Belczynski \&
Kalogera 2001, hereafter BK01).  First, we have replaced the
approximate prescription suggested by Bethe \& Brown (1998) for the
hyper-critical accretion during CE phases, with a newly derived
numerical solution (see Appendix). Second, we allow for hyper-critical
CE evolution of low-mass helium stars with compact objects. In the
studies presented in BK01, we allowed binaries with low-mass helium
giants to evolve through DCE and standard SCE, but we had assumed that
CE events of helium giants with compact objects lead to mergers, and
possibly a gamma-ray burst (e.g., Fryer et al.\ 1999).  However, due to
the small mass of helium giant envelope at the onset of CE event ($\sim
1-1.5 M_\odot$), we find that these systems survive the CE events, and
form very tight NS-NS binaries (see also footnote in \S\,2.1.2).

Double neutron stars are formed in various ways through more than 14
different evolutionary channels identified in Table~\ref{channels}.
However, the entire population of coalescing NS-NS systems, may be
divided into three subgroups.

{\em Group I.} This subpopulation consists of non-recycled NS-NS
systems, first identified by BK01. These are systems in which none of
the two NS ever had a chance of getting recycled through accretion.  
Our current results for the predicted formation rates and properties of
the non-recycled NS-NS systems, have not been affected by the two
improvements discussed above.  As shown in Table~\ref{channels}, these
recently identified non-recycled NS-NS systems are formed via the
NSNS:09, NSNS:11 and NSNS:12 channels, which involve DCE of two low-mass
helium giants, which were already allowed in the earlier version of {\em
StarTrack}.

The unique qualitative characteristic of this NS-NS formation path is
that both NS have avoided recycling. The NS progenitors have lost both their
hydrogen and helium envelopes prior to the two supernovae, so no accretion
from winds or Roche-lobe overflow is possible after NS formation.
Consequently, these systems are detectable as radio pulsars only for a
time ($\sim 10^6$\,yr) much shorter than recycled NS-NS pulsar lifetimes
($\sim 10^8-10^{10}$\,yr in the observed sample). Such short lifetimes are
of course consistent with the number of NS-NS binaries detected so far
and the absence of any {\em non-recycled} pulsars among them.

We note that the identification of the formation path for non-recycled
NS-NS binaries stems entirely from accounting for the evolution of helium
stars and for the possibility of double CE phases, both of which have
typically been ignored in previous calculations (with the exception of
Fryer et al.\ 1999, where, however, such events were assumed to lead to 
mergers).

{\em Group II.} This subpopulation consists of tight, short lived
binaries with one recycled pulsar. Their merger times are typically
$\sim$\ 1 Myr or even smaller (see \S\,3.4.5). As shown in
Table~\ref{channels}, these new dominant NS-NS systems are formed via
the NSNS:01--08, NSNS:10 and NSNS:13 channels, with the common
characteristic that the {\em last} binary interaction is a
hyper-critical CE of a low-mass helium giant and the first-born NS.

In Belczynski, Bulik, \& Kalogera 2002a we describe in detail the
formation of a typical NS-NS binary of group II. The most channel
identified as the most efficient for NS-NS formation (NSNS:01)
corresponds to the ``standard'' channel of Fryer et al.\ (1999).  The
only difference is an extra CE event which originates from allowing for
helium star evolution and without a priori assumptions about the CE
outcome. The second most dominant channel, involving two consecutive MT
episodes and then two SN explosions, closely resembles our channels:
NSNS:02, NSNS:04, NSNS:06, NSNS:10, NSNS:11, NSNS:12.  The only
difference again remains an extra MT episode from evolved,
Roche-lobe-filling helium stars.

The most dramatic effect of the binary evolution updates is reflected in
the existence of a whole new population of coalescing NS-NS stars formed
in the Group II. In our standard model these channels contribute 50\% of
the DCO population, and their common characteristic is that the {\em
last} binary interaction is a hyper-critical CE of a low-mass helium
giant and the first-born NS.  It turns out that the majority of these
systems survive the HCE event and form tight NS-NS binaries. Had we not
taken into account the radial expansion of low-mass helium-rich giants,
the progenitors of this dominant NS-NS population would have evolved
without any further MT. Most of them would have still formed NS-NS
systems, however, not as tight as after this last CE episode. We have
actually examined this alternative and found that about half of them
would have formed binaries with merger lifetimes longer than 10$^{10}$\
yr. Once again, we see the importance of helium star evolution on DCO
population synthesis.

{\em Group III.} This subpopulation consists of all the other NS-NS
systems (belonging neither to Group I nor II) formed, through more or
less classical channels (Bhattacharya \& van den Heuvel 1991). The
formation path denoted NSNS:14 corresponds to what is usually considered
to be the ``standard'' NS-NS formation channel (Bhattacharya \& van den
Heuvel 1991). Since we account for hyper-critical accretion in CE, the
formation rate is decreased because some NS (but not all, as assumed by
Portegies Zwart \& Yungelson 1998 and by Fryer et al.\ 1999)  collapse
to BH. Furthermore our treatment of the hyper-critical accretion
typically leads to tighter post-CE systems, causing more binaries to
merge in CE events, and thus decreases the number of possible NS-NS
progenitors.

Group II strongly dominates the population of coalescing NS-NS systems
(81\%, for standard model calculation) over group III (11\%) and I
(8\%). In general this characteristic is preserved in all, except a few
extreme evolutionary models. The discussion of the dependence of the
formation rates of the various NS-NS subpopulations on population
synthesis model assumptions will be presented in Belczynski et al.\
2002a.

\subsubsection{Populations of Black Hole Binaries} 

In general, BH-NS and BH-BH binaries are formed through just a few
distinct channels, with a moderate number of MT events (2--3), in
contrast to our findings for NS-NS systems.

Helium star evolution, radial expansion and CE phases are much less
important for the formation of BH-NS and BH-BH binaries.  The reason is
that for most of these progenitors the first-born compact object is
massive enough that even when helium stars evolve to the giant branch,
they do not expand to large radii nor they lead to possible CE evolution
(see MT criteria in \S\,2.2.4 and channel BHNS:03 in
Table~\ref{channels}).  Instead these DCO form most efficiently through
channels that closely resemble those NS-NS conventionally thought to be
``standard'' (Bhattacharya \& van den Heuvel 1991; Fryer et al.\ 1999):
evolution is initiated with a phase of non-conservative mass transfer
and followed either by a CE phase or the formation of the first compact
object (see BHNS:01, BHNS:02, and BHBH:01, BHBH:02).

\subsection{Physical Properties of Double Compact Objects}

\subsubsection{Double Compact Objects Component Masses}

\begin{figure*}[t]
\centerline{ \psfig{file=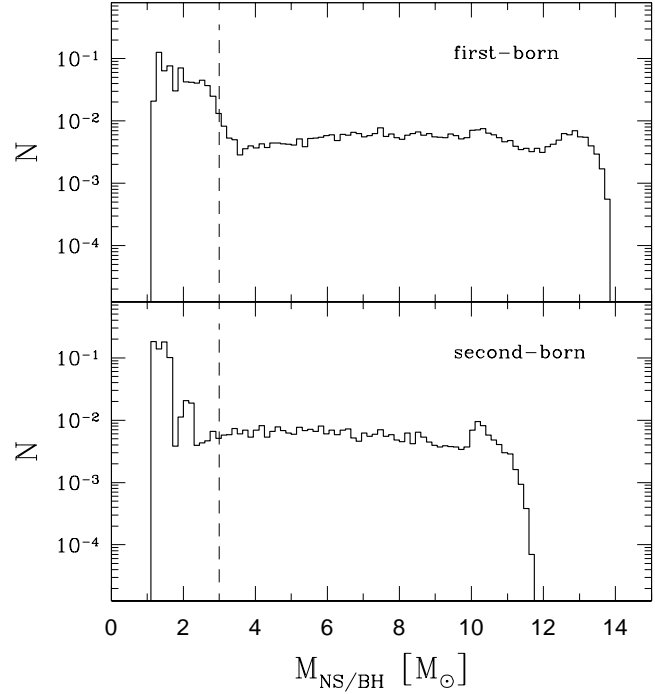,width=0.5\textwidth} }
\caption{
Normalized distributions of the compact object masses in coalescing
NS-NS, BH-NS and BH-BH systems for the standard model. We show the mass
distributions for first- and second-born compact objects in the top and
bottom panel, respectively.  The distributions are calculated using a
total number of 28655 coalescing DCO formed out of $10^7$ 
primordial binaries. Note that the standard model
choice of the maximum NS mass is: $M_{\rm max}^{\rm NS}=3\,M_\odot$ (marked with 
the dashed line).}
\label{fig2}
\end{figure*}

{\em Standard Model.} In Figure~\ref{fig2}, we present the mass
distributions of DCO components\footnote{Mass distributions of 
compact objects with white dwarf companions as well as single compact 
objects (formed both from single star progenitors and from components 
of disrupted binaries) are presented in 
Belczynski, Bulik \& Kluzniak 2002b.} 
for our standard model, with first- and
second-born compact objects in the top and bottom panel, respectively.

The shape of the calculated mass distributions can be understood as a
convolution of the initial-remnant mass relation for single stars (top
panel of Figure~\ref{fig1}) with the adopted initial stellar mass
function (see eq.~\ref{imf}). Deviations from such a convolution reflect
the effects of binary evolution on the compact object masses.

First, we discuss the mass distribution of the second-born compact
objects, as they are less affected by binary evolution.  This
distribution starts at the lowest possible NS mass (allowed by our
single star models) of $1.2 {\rm M}_\odot$, rises sharply with a peak at
$\approx 1.4 {\rm M}_\odot$ and then declines down to $\approx 3 {\rm
M}_\odot$ (the assumed $M_{\rm max}^{\rm NS}$ for our standard model). This
strong peak at low masses is the result of a rather weak dependence of
NS mass on ZAMS mass (Figure~\ref{fig1}) combined with the steep IMF.
Following the initial peak, the distribution becomes flat in the range
$\approx 3-10 {\rm M}_\odot$. This flattening is caused by the balanced
effects of a slightly rising initial-remnant mass relation and a
declining IMF.

At $\sim 10.5 {\rm M}_\odot$ the distribution peaks again, and then
rapidly declines.  This final peak corresponds to the saturation of BH
masses at $\sim 10.5 {\rm M}_\odot$ for a wide range of ZAMS masses in the
initial-remnant mass relation.

For the model described here, the maximum mass of the second-born compact
objects is $\approx 12 {\rm M}_\odot$, in excess of the maximum mass of
$\approx 11 {\rm M}_\odot$ of compact objects formed in single star
evolution (Figure~\ref{fig1}).  Such slightly more massive compact objects are
allowed only because of MT episodes involving their progenitors. In some
cases they can accrete enough material to significantly increase the core
mass (rejuvenation through accretion), and form more massive remnants.

The mass distribution of first-born compact objects shows more prominently
the effects of binary evolution, since the compact objects themselves and
not just their progenitors might have been affected by mass transfer
events.  The three basic qualitative features are still present, with an
initial peak, a subsequent flattening, and a smaller final peak. However,
there are two easily identified differences between the two distributions.  
First, the initial peak is lower and broader, and second, the peak at high
masses shifts to $\approx 13 {\rm M}_\odot$, and the maximum compact
object mass increases to $\approx 14 {\rm M}_\odot$. Both effects are due
to accretion onto the first-born compact objects from their non-degenerate
companions.

Our standard-model mass distribution of compact objects in coalescing
double compact objects peaks at $\sim 1.4 {\rm M}_\odot$ for NS and then
extends up to $\sim 14 {\rm M}_\odot$ for BH. This agrees well with the
observed NS masses in binary stars:  $1.1-1.6 {\rm M}_\odot$ (e.g.
Thorsett \& Chakrabarty 1999) and also with estimated masses of black hole
binary candidates:  $\sim 3-20 {\rm M}_\odot$ (for references see
\S\,2.1.3).  It is important to note that the maximum BH mass of $\sim 14
{\rm M}_\odot$ can be increased up to $\simeq 20-25 {\rm M}_\odot$. This
can be achieved either by decreasing the wind mass loss rate by factor of
2 (see the bottom panel of Figure~\ref{fig1}) and/or by increasing the 
amount of material accreted by stars in non-conservative MT events.

We can also compare results of our population synthesis calculations to
the theoretical compact object mass distributions presented by Fryer \&
Kalogera (2001, hereafter FK01; their Figures 7 and 8).  We find some striking
similarities. The single star initial-remnant mass relations are very
similar as expected given that our calculation of compact object masses
are based on the same core-collapse hydrodynamical calculations (Fryer
1999).  FK01 find that in the range of compact object formation 80\% of
their single star remnants are NS, and the rest are BH. We find that for
the standard model of single star evolution 81\% of remnants are NS, and
19\% are BH.

More importantly, the {\em shape} of the FK01 CO mass distribution
formed in the presence of stellar winds and binary companions, resembles
our distribution, even though FK01 did not use population synthesis.
Apart from the qualitative similarities, we also note that our results
confirm the findings of FK01 for a continuous mass distributions and the
absence of a gap or isolated narrow spike at about $7 {\rm M}_\odot$,
claimed by Bailyn et al. (1998). Recent observational results also point
in the direction of a broader range of BH masses (Froning \& Robinson
2001; Orosz et al. 2001).  However, there are still two significant
differences between our results and those of FK01. One is related to the
maximum BH mass (about $10 {\rm M}_\odot$ in FK01 compared to our $14
{\rm M}_\odot$) and the other to the contribution of BH to the compact
object population. Both are linked to a number of binary interactions
that are not taken into account by FK01 and have the effect of
increasing the masses of compact objects. In our standard-model
calculations we find that, in double compact objects, NS and BH
represent 65\% and 35\% of the compact objects, respectively, because of
accretion effects. In contrast, for the case in FK01 where effects of
stellar winds and binary companions, the contribution of BH actually
decreases compared to single star calculations.

{\em Parameter Study.} We have explored a large number of population
models and have found that the main qualitative features of the mass
distributions remain unaffected:  the low-mass peak from NS formation
and the flat form of the distributions for a wide range of BH masses are
present in many different models (e.g., B1, B12, F2, J). Accretion
effects and the less prominent peak at higher masses are also clearly
seen in almost all of the models.

Nevertheless some dramatic quantitative differences are revealed by a 
few models. In the case that all wind mass-loss rates are increased by a
factor of 2 (G2), the maximum compact object mass turns out to be only
$\simeq 3.5\,M_\odot$ and therefore such a case appears to be highly
unlikely. On the other hand, the model where all wind mass-loss rates are
decreased by a factor of 2 (G1) allows for BH masses as high as $\simeq
21\,M_\odot$. 

\begin{figure*}[t]
\centerline{ \psfig{file=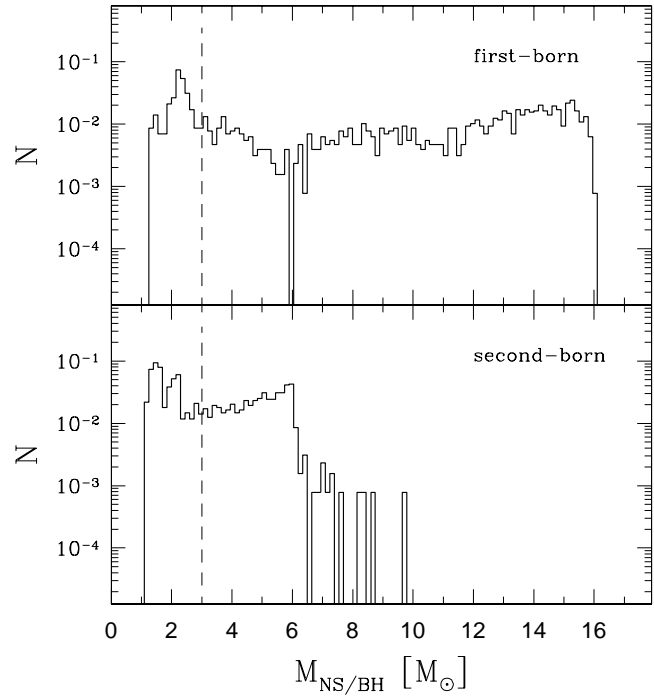,width=0.5\textwidth} }
\caption{
Normalized distributions of the compact object masses in coalescing
NS-NS, BH-NS and BH-BH systems for model E1 with very low CE efficiency:
$\alpha_{\rm CE}\times\lambda=0.1$. We show the mass distributions for
first- and second-born compact objects in the top and bottom panel,
respectively. The distributions are calculated using a total number of 1283
coalescing DCO formed out of $3 \times 10^6$ primordial binaries. 
This case of very low CE efficiency results in very low DCO rates and 
relatively small-number statistics.
Note that the standard model choice of the maximum NS mass is: 
$M_{\rm max}^{\rm NS}=3\,M_\odot$ (marked with the dashed line).
}
\label{fig3}
\end{figure*}

Another class of models that show significant quantitative differences
from the standard model are those with varying CE efficiencies,
especially those with low values. The results for model E1 ($\alpha_{\rm
CE}\times\lambda=0.1$) are shown in Figure~\ref{fig3}.  Although the
basic shape of the distributions resembles that of the standard model,
the second-born compact objects span a much narrower range in mass:
$1.15-9.7 \,M_\odot$ (with most of compact objects formed with mass
smaller than $6 \,M_\odot$).  Low CE efficiency discriminate against
tight pre-CE systems, cause a large number of DCO progenitors to merge,
and therefore DCO rates to decrease significantly. In particular, the
effects appear to be most prominent for BH-BH binaries where massive
second-born BH are essentially eliminated. As shown in \S\,3.3 and
Table~\ref{channels} most BH-BH progenitors experience first a
non-conservative MT and then a CE episode. For the case of low CE
efficiencies, CE survival is favored for rather wide pre-CE systems with
donors (stars that initiate the CE episode and progenitors of the
second-born compact objects) with low envelope masses. In these cases CE
ejection is facilitated, but at the same time lower mass progenitors are
favored, leading to a restricted mass range for these second-born
compact objects. At the same time the bias in favor of wide pre-CE
systems translates to a bias in favor of progenitors where the first MT
episode is initiated late in the evolution of the primary, when the
stellar radii are larger and most of the stellar envelope has burnt into
a high mass core.  This bias clearly favors the formation of higher mass
first-born compact objects, leading overall to BH-BH binaries with
rather extreme mass ratios.

Given the sensitivity of the DCO mass distribution to the CE
efficiencies, it may be possible in the future to discriminate between
models based on mass measurements of BH-BH binaries from
gravitational-wave detections and data analysis.

\subsubsection{Double Compact Objects Orbital Separations}

\begin{figure*}[t]
\centerline{ \psfig{file=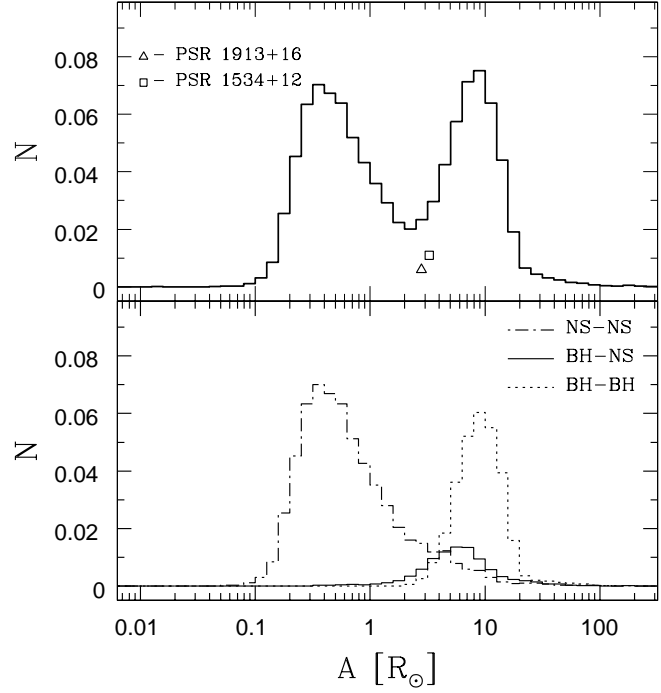,width=0.5\textwidth} }
\caption{
Normalized distributions of orbital separations of 
coalescing DCO at the time of their formation, for our standard model, 
for all DCO (top) and for the three classes separately (bottom).
For comparison, we also show the observed orbital separations of the 
2 coalescing NS-NS systems found in the Galactic field, PSR 1913+16
and PSR 1534+12.}
\label{fig4}
\end{figure*}

{\em Standard Model.} The distribution of DCO over orbital separations 
at the time of their formation is shown in Figure~\ref{fig4}, for our 
standard model.

The distribution for the whole population (top panel) covers a wide
range of values $\sim 0.1-100 {\rm R}_\odot$ and is characterized by two
distinct peaks. It is evident from the bottom panel that the peak at
shorter separations is due to NS-NS binaries and the one at wider is due
to BH binaries.  Examination of the NS-NS distribution indicates that
the majority of NS-NS systems, are formed with very small orbital
separations ($< 1 {\rm R}_\odot$), but there is a long tail stretching
to much larger orbital separations (even beyond $10 {\rm R}_\odot$).  
This result is related to our discussion of the main NS-NS formation
channels (\S\,3.3). The NS-NS population consists of three distinctive
groups (see \S\,3.3.1), which shape the orbital separation distribution.  
``Classical'' NS-NS systems (group III) that contain recycled pulsars
have wide separations, mostly in the range $\sim 1-10 {\rm R}_\odot$
(where the two observed systems lie).  However, NS-NS binaries in which
none of the NS ever had a chance of getting recycled (group I) {\em and}
very tight NS-NS binaries formed via the newly identified channels
(group II) actually dominate the produced NS-NS population and produce a
strong peak at much lower orbital separations. These differences in
orbital separations affect the derived merger times and have important
implications for the detection of such systems, which we discuss in
\S\,3.5.2.

Typical BH binaries have much wider separations than NS-NS systems
(Figure~\ref{fig4}). This is a result of the lower (or zero) SN kicks
imparted to BH which allows the survival of wider progenitors, typical
of the more massive binaries required for the formation of the BH
binaries. Note that BH-NS binaries tend to have intermediate size
orbits, affected in part by the high kicks imparted to one of the two
compact objects.

{\em Parameter Study.} The double-peaked shape of the orbital separation
distributions found in the standard model persists in a robust way in
the majority of the models in our parameter study. The full range of
separation values also appears to be robust (0.1--100 R$_\odot$).  
However, the relative height of the two peaks varies significantly from
model to model, primarily because of the varying contribution of
different types of systems to the DCO population. The reasons and a
discussion of these variations are given in \S\,3.5.1.

Here, we identify a couple of extreme cases: In the models with zero kick
velocities (B1) or increased wind mass-loss rates (G2), BH binaries have a
very small formation rate relative to NS-NS systems(see also Portegies
Zwart \& Yungelson 1998; Fryer et al.\ 1999), and therefore the NS-NS peak
at shorter orbital separations strongly dominates. In contrast, models
with low CE efficiencies hamper the formation of DCO in tight orbits and
the peak due to BH binaries strongly dominates. In model H2, as expected,
the peak at low separations for NS-NS disappears altogether, since no
helium-star CE episodes are allowed.

\subsubsection{Double Compact Objects Eccentricities}

\begin{figure*}[t]
\centerline{ \psfig{file=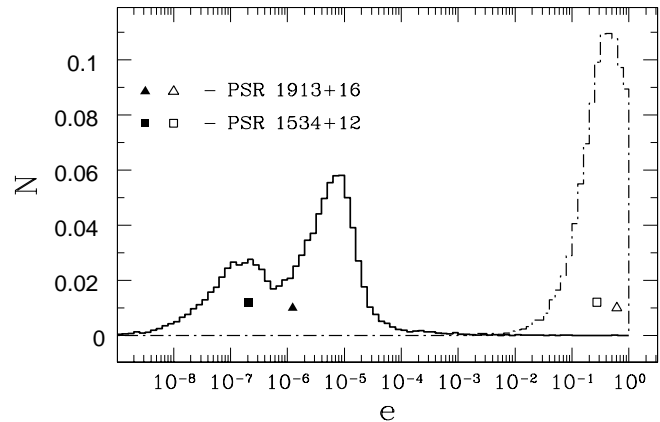,width=0.5\textwidth} }
\caption{
Normalized eccentricity distributions of coalescing DCO for our 
standard model at the time of their formation (dot-dashed line) and at
the time the orbit decayed so that the corresponding gravitational-wave
frequency is about 40\,Hz, i.e., the system is entering the LIGO I band
(solid line).  For comparison we also show the two observed short-period
NS-NS systems: their current (open symbols) eccentricities and the 
eccentricities as they enter the LIGO I band (filled symbols).}
\label{fig5}
\end{figure*}

{\em Standard Model.} In Figure~\ref{fig5}, we present eccentricity
distributions of coalescing compact object binaries for our standard
model. The values at the time of DCO formation are quite high, $e>0.1$
for the majority of the population, and of course originate from the
fact that the last stage prior to DCO formation typically involves a SN
explosion and asymmetric kicks. The fraction of DCO formed in circular
orbits is rather small, and occurs when the second compact object is
formed through a direct collapse to a BH (BH-BH: $\simeq 2.4\%$ and
BH-NS:$\simeq 0.1\%$).  Portegies Zwart \& Yungelson (1998) and Fryer et
al.\ (1999) have also found that NS-NS population is born with high
eccentricities, which is in good agreement with our results.

Apart from the eccentricities at formation, we also examine the
distribution at later stages when gravitational radiation has caused the
binary orbit to shrink and the gravitational-wave
frequency has increased to $\simeq 40$\,Hz, the lower end of the LIGO I
band. So far DCO have been assumed to be circular for purposes of
gravitational-wave detection and data analysis. We examine whether our
model population satisfies this assumption due to angular momentum 
losses
and circularization. We use the dependence of orbital separation on
eccentricity due to gravitational-wave emission as derived by
Peters (1964):
 \begin{equation}
 A(e)= {c_0 e^{12/19} \over (1-e^2)} \left[1+{121 \over 304} e^2 
\right]^{870/2299}, 
 \end{equation}
 where $c_0$ is given in equation~\ref{mergtim} and depends on the
orbital separations and eccentricities at DCO formation. We use our
results for the masses to calculate first the orbital separation at
40\,Hz and the corresponding frequency. The resulting distribution is
shown in Figure~\ref{fig5} (solid line). It is clear that by the time
these coalescing DCO enter the LIGO I band, the eccentricities are small
enough ($e<0.0001$) that the assumption of circular orbits is well
justified.

It is interesting to note that, although the eccentricity distribution 
at birth is single-peaked, the distribution at later stages becomes 
double-peaked. This is a direct result of the double-peaked form of the 
orbital separation distribution. In more detail, the peak at lowest
eccentricities is populated by the heavier BH binaries and the peak at 
somewhat higher eccentricities by NS-NS binaries. The distinction is 
related to the weaker effects of gravitational-wave emission for the 
less massive systems.

{\em Parameter Study.} The shape and range of typical values for the
orbital eccentricities appear to be quite robust. Differences becomes
noticeable only in models of varying kick-magnitude distributions.  As
expected, the typical eccentricities at formation decrease with a
decreasing average kick magnitude.

The variations in the shape and position of the eccentricity peaks for
later stages are more prominent and closely follow the corresponding
variations in the distributions over orbital separations. Nevertheless,
the eccentricities at 40\,Hz remain below $10^{-3}$ for all the models
we have examined.

\subsubsection{Double Compact Objects Center of Mass Velocities}

\begin{figure*}[t]
\centerline{ \psfig{file=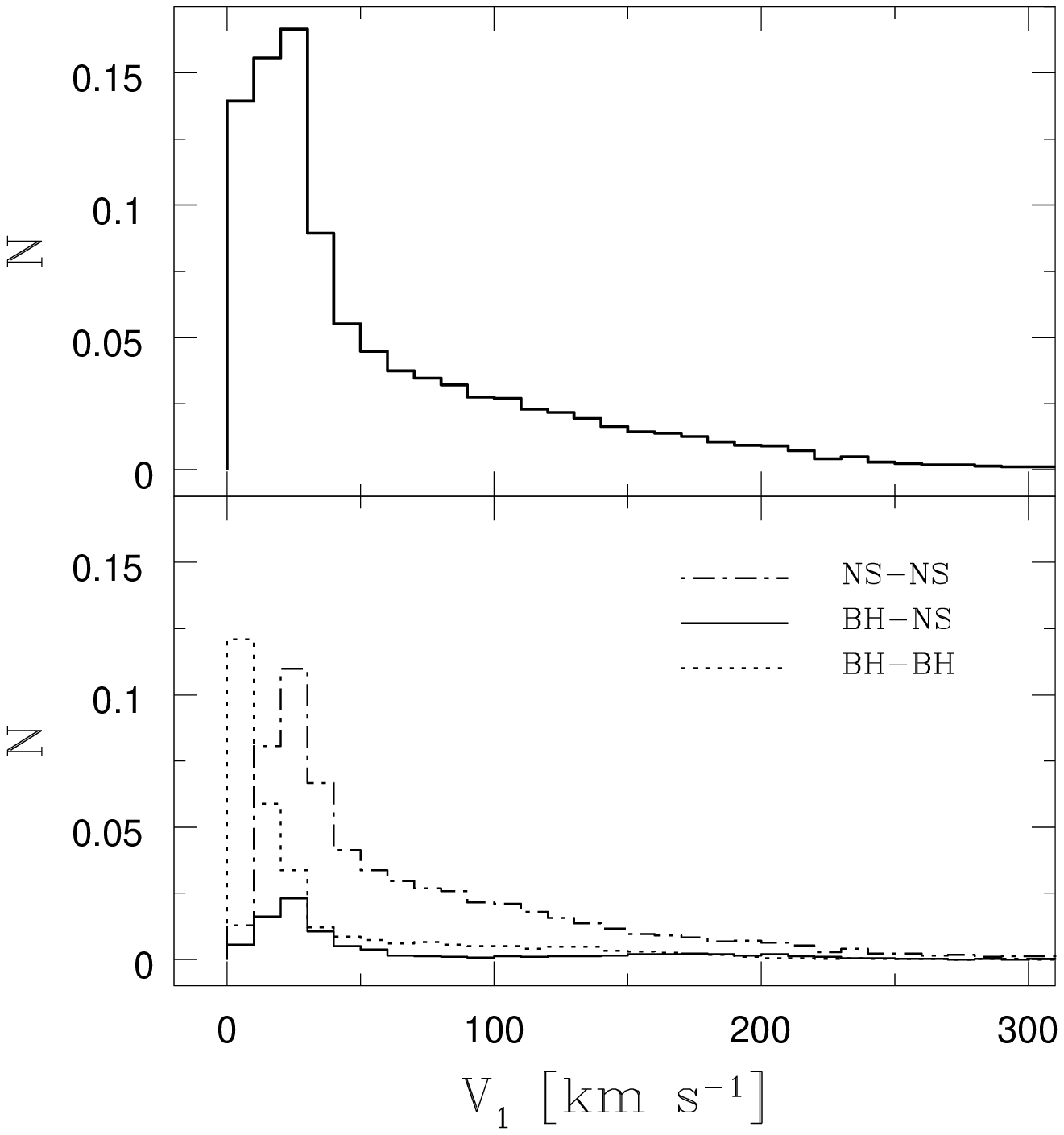,width=0.5\textwidth} } 
\caption{
Normalized distributions of DCO systemic velocities acquired after the
first core-collapse event, for our standard model, for the whole 
population (top) and for each DCO class separately (bottom).}  
\label{fig6}
\end{figure*}

\begin{figure*}[t]
\centerline{ \psfig{file=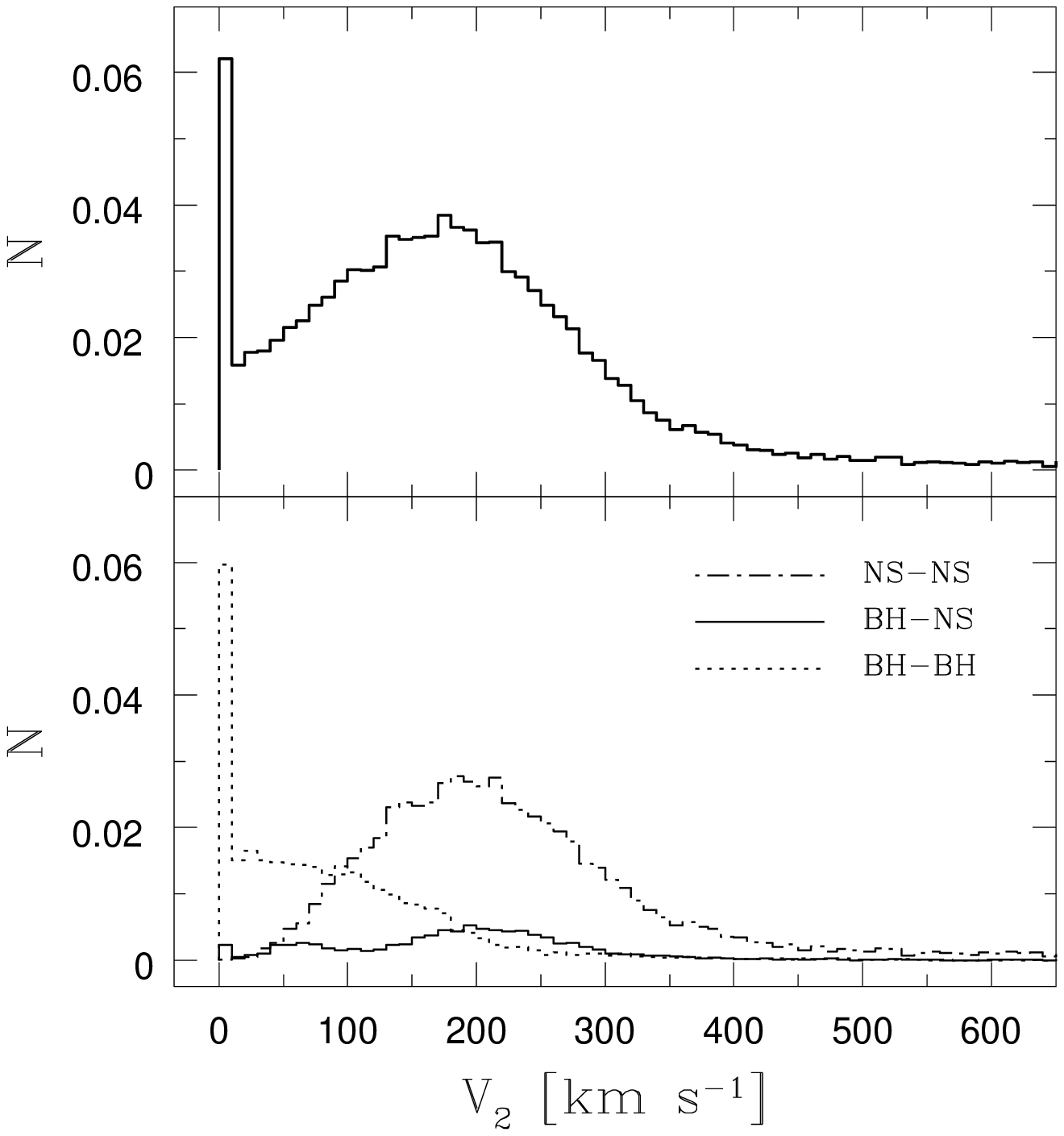,width=0.5\textwidth} }
\caption{
Normalized distributions of DCO systemic velocities acquired after the
second core-collapse event, for our standard model, for the whole 
population (top) and for each DCO class separately (bottom).} 
\label{fig7}
\end{figure*}

{\em Standard Model.} Binaries are expected to acquire systemic
(center-of-mass) velocities after core-collapse events because of the
combined effects of mass loss and SN kicks imparted to the compact
remnants. Understanding of these velocities is crucial in studies of
dynamical evolution of these populations, and their kinematic properties
and spatial distributions with respect to host galaxies.  The
distributions of center-of-mass velocities after the first and second
core-collapse events are shown in Figures~\ref{fig6} and ~\ref{fig7},
respectively.  Top panels on both figures show the distribution for the
whole DCO population, and bottom panels show the distributions
separately for each of the DCO groups.

It is evident that there are major differences between the two plots.
The distribution after the first core-collapse event (V$_{\rm 1}$) shows
a well defined narrow peak at low velocities ($\sim 25 {\rm 
km\,s}^{-1}$)
and a slowly decaying tail at higher velocities.  After the formation of
the second compact object velocities (V$_{\rm 2}$) populate a much
broader peak positioned at high magnitudes ($\sim 200 {\rm km\,s}^{-1}$)
with a velocity tail reaching to very high velocities ($> 600 {\rm km
s}^{-1}$). Moreover, in these distributions a narrow spike 
at $0 {\rm km\,s}^{-1}$ is also identified.

All these characteristics can be naturally explained by the different
binary orbital properties of binaries at the time of the first and the
second compact object formation. For the majority of DCO progenitors,
the first core-collapse event happens when the binary orbit is still
wide, accommodating two massive stars and is affected primarily by mass
loss through winds and non-conservative mass transfer events. The
typical orbital velocities in these pre-collapse systems are $\sim
10-50$\,km\,s$^{-1}$ and it is mostly these orbital velocities that
determine the typical post-collapse systemic velocities of bound systems
(Kalogera 1996). For Maxwellian kick magnitude distributions an upper
and lower limit on the systemic velocity can be derived analytically and
it is found to be independent of the average kick magnitude (depends
only on the stellar masses and the pre-collapse relative orbital
velocity; see Kalogera 1996). As a secondary effect, within the range of
values defined by these two limits, systems tend to acquire higher
systemic velocities when higher kicks are imparted. A smaller fraction
of DCO progenitors experience a CE phase prior to the first
core-collapse event, and have much tighter orbits (orbital velocities of
order $\sim 100$\,km\,s$^{-1}$), leading to higher systemic velocities.
The more subtle influence of the average kick magnitude is evident when
looking at the distributions for each DCO class, where NS-NS progenitors
are slightly shifted to higher and BH binaries to lower systemic
velocities (NS-NS progenitors tend to be lighter than BH-BH
progenitors).

At the time of the second SN explosion, the population of compact object
binary progenitors is dominated by tight binaries with relative orbital
velocities in the low hundreds. The result is a strong, broad peak at
$100-300$\,km\,s$^{-1}$ with a tail to higher velocities
(Figure~\ref{fig7}). Systems with lower systemic velocities do form, but
the majority of them are so wide that their merger times exceed the
Hubble time.  Another noticeable difference from Figure~\ref{fig6} is
that the BH-NS population now closely follows the NS-NS population.  
This is understood as both the pre-collapse orbital characteristics and
the typical kick magnitudes are similar for these two populations at the
second compact object formation, since in the majority of BH-NS systems
the NS is formed second. It is evident that typically BH-BH binaries
acquire the lowest systemic velocities.  This is a result of three
combined effects: they are formed from wider progenitors (\S\,3.4.2),
they are heavier systems, and some BH are formed via direct collapse and
hence do not acquire any kick. Strikingly, in both Figures~\ref{fig6}
and ~\ref{fig7}, there is a prominent spike at zero systemic velocity,
which is populated by BH-BH systems where both BH were formed via
direct, assumed to be symmetric collapse.  We note that, although these
systems tend to be wider than NS-NS, they still merge within a Hubble
time because of the higher masses, hence stronger gravitational radiation
involved.

It is useful to note one other implication of the small systemic
velocities after the first core-collapse event. These combined with the
fact that the time between the two collapse events is much shorter than
typical coalescence times, implies that it is the systemic velocities
after the second collapse (along with the merger times) that determine
the spatial distributions of DCO merger sites with respect to host
galaxies.

Portegies Zwart \& Yungelson (1998) presented results on NS-NS systemic
velocities, and Fryer et al.\ (1999) calculated systemic velocities for
both NS-NS and BH-NS binaries. Comparison with our results shows overall
good agreement; both Portegies Zwart \& Yungelson (1998) and Fryer et
al.\ (1999) found that most systems acquire systemic velocities of the
order of $\sim 200-300$\ km\,s$^{-1}$, with some binaries being
accelerated to much higher velocities ($\sim 500-1000$\ km\,s$^{-1}$).

{\em Parameter Study.} The distribution of systemic velocities after the
first core-collapse event is found to be very robust in our parameter
study.  The most clear, but not big, changes in the distribution shape
are visible in models which affect the pre-SN orbital separations, e.g.
in models E1 and F2. For these models, the distribution starts with an
initial peak, as for the standard model, however the high velocity tail
is depleted at velocities below 100--150 km\,s$^{-1}$. All of DCO
progenitors experience either CE phase or MT event prior to the first SN
explosion. Decreasing the CE efficiency (model E1) or increasing the
fraction of mass accreted by companions in MT events (model F2) lead to
a decrease in post-CE or post-MT separations. Therefore, at the time of
the first SN explosion, progenitors have higher orbital velocities, and
DCO progenitors in models E1 and F2 end up with higher systemic
velocities than in the standard model, depleting the distribution of low
velocity values. Some small changes are also seen with kick magnitude
variations.  For very small kicks, the high velocity tail of the
standard model distribution disappears, whereas, for very high kicks,
many more systems are disrupted, but most of the ones that survive the
collapse, end up with somewhat higher systemic velocities. As a result,
the low velocity peak of the distribution is depleted and the tail
extends to higher velocities.

Similarly, the qualitative features of the distributions after the second
compact object formation are robust, with an exception for the spike at
zero velocity, the presence of which depends on the contribution of BH-BH
binaries to the DCO population (e.g., models B1, B6, B13, D2, F2, J, and
the extreme case of G2 where essentially no BH-BH binaries are formed).
For a few models, we find some significant differences, but are all
explained by the relative contribution of the various DCO groups.  For
example, for very high kicks (model B12) and in the absence of helium-star
CE phases (model H2), the formation rate of coalescing NS-NS is so small
that the velocity distribution is dominated by BH-BH binaries, and in
particular those which do not receive a kick (i.e., the zero-velocity
spike becomes most dominant). The few surviving NS binaries populate a
much broader range of systemic velocities with low normalization and a
roughly flat tail out to hundreds of km\,s$^{-1}$.

\subsubsection{Double Compact Objects Merger Times}

\begin{figure*}[t]
\centerline{ \psfig{file=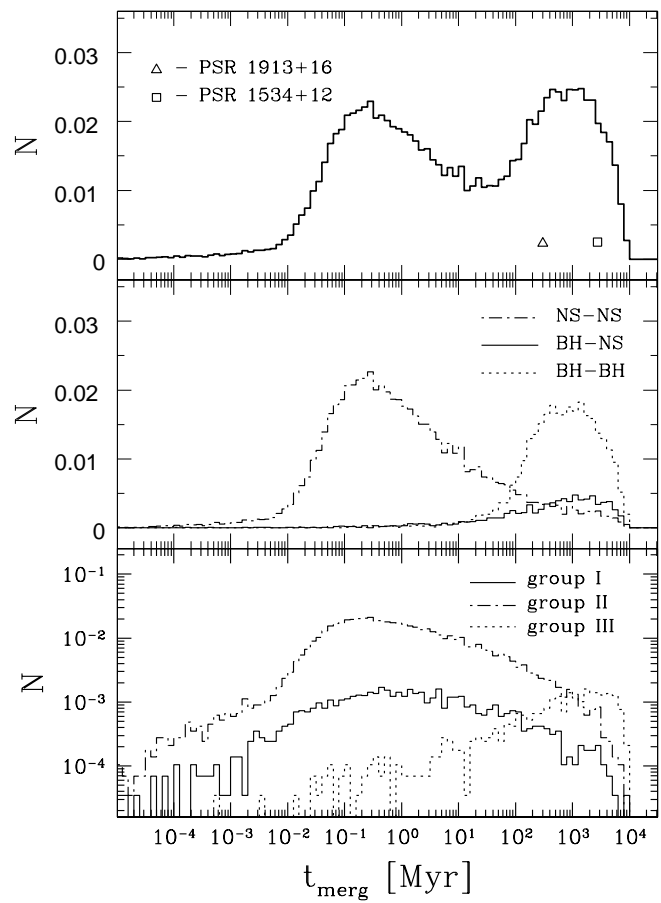,width=0.5\textwidth} }
\caption{
Normalized merger time distributions shown for our standard model, for
the whole DCO population (top), for each of the DCO classes (middle), and
for each of the three major NS-NS groups (bottom). 
The merger of the two observed systems are also shown in the
top panel.}
\label{fig8}
\end{figure*}

{\em Standard Model.} The distributions of merger times (see
eq.~\ref{mergtim}) of coalescing DCO are shown in Figure~\ref{fig8}, for
our standard model. The double-peaked form is due to the two main
populations of NS-NS systems and BH binaries and their double-peaked
distributions of orbital separations. As already mentioned the NS-NS
population is dominated by tight binaries formed through channels
involving MT episodes from helium stars. These tight orbits imply very
short merger times with a peak at $\sim 0.3$\,Myr. Despite their higher
masses BH-NS and BH-BH binaries are found to have merger times typical
of $\sim 1$\,Gyr, driven by their wider orbits and the stronger
dependence of merger time on separation.  Merger times also decrease
with increasing initial eccentricities, leading to a stronger separation
of the two peaks, since NS-NS binaries tend to have not only tighter
orbits, but also higher eccentricities.

In the bottom panel of Figure~\ref{fig8}, we present the merger times of
three different populations of coalescing NS-NS binaries (see
\S\,3.3.1):  group I of tight, non-recycled NS-NS binaries; group II of
tight binaries formed through the newly identified channels; and group
III of NS-NS formed through ``classical'' channels.  It is evident that
the large relative contribution of the tight NS-NS binaries drives the
typical NS-NS merger times down to values close or below 1\,Myr, whereas
the ``classical'' NS-NS have merger times typical of the BH binaries.
The identification of these short-lived binaries has important
implications for the detectability of coalescing NS-NS binaries (see
\S\,3.5.2).

Both, Portegies Zwart \& Yungelson (1998) and Fryer et al.\ (1999),
presented merger time distributions of NS-NS binaries. They obtained
typical times much longer than ours, close to 100--1000 Myr. This
discrepancy originates from the newly identified short-lived NS-NS
systems.  However, our classical population of NS-NS binaries have, as
expected, merger times comparable to these found in these two earlier 
studies.

\begin{figure*}[t]
\centerline{ \psfig{file=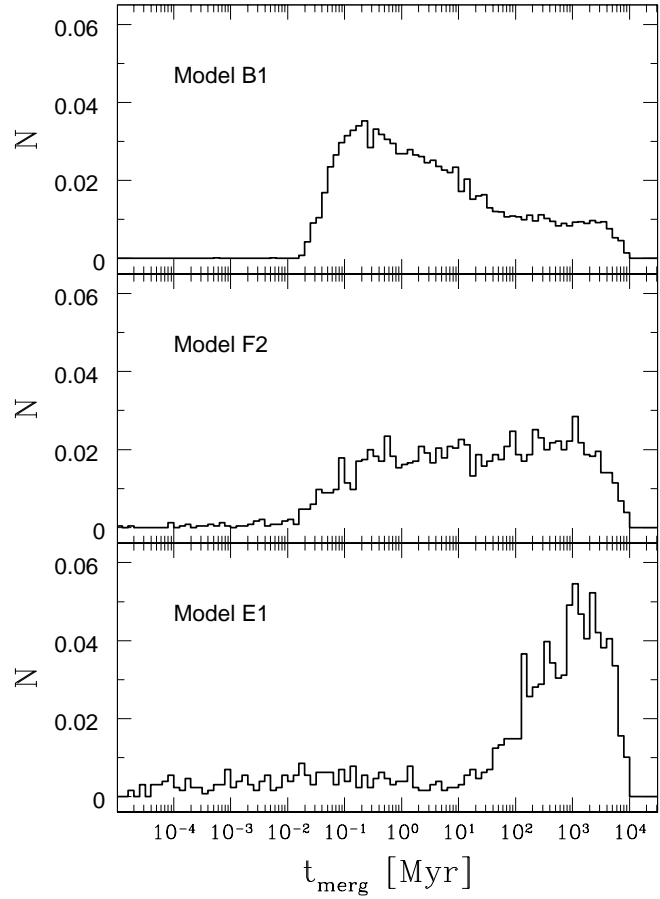,width=0.5\textwidth} }
\caption{
Normalized merger time distributions of coalescing DCO for models B1 (top
panel), F2 (middle panel) and E1 (bottom panel). 
The distributions are normalized to the total number of coalescing DCO: 
10841, 2352, 1283 for models B1, F2 and E1, respectively; 
formed out of $10^6$\ primordial binaries for model B1 and F2, 
and out of $3 \times 10^6$\ binaries for model E1.}
\label{fig9}
\end{figure*}

{\em Parameter Study.} Unlike all other binary properties we have
examined, the qualitative characteristics of the merger-time
distributions appear to be rather sensitive to a number of model
parameters. This is understood in terms of the strong dependence of
merger times on initial separations and eccentricities, and, although
their distributions change only slightly with model parameters, the
corresponding change of merger time distributions is quite dramatic. In
Figure~\ref{fig9}, we show three of the most different DCO merger time
distributions from models with somewhat extreme assumptions.

In the top panel of Figure~\ref{fig9} (model with zero kicks, B1), the DCO
population is dominated by the NS-NS binaries, causing at first a sharp
rise followed by a slow decline at longer merger times (compare to the
middle panel of Figure~\ref{fig7}). The flattening of the distribution
at $10^2-10^4$\ Myr is due to the small contribution of BH-NS and BH-BH
systems to the DCO population in this model. In the model shown in the
middle panel (model F2), all dynamically stable MT episodes are assumed to 
be conservative. The distribution is rather flat over a wide range of
merger times and this is mainly a result of the assumption of
conservative MT and the wider orbits that form as a consequence.  In the
bottom panel, we show results from a model with low CE efficiency (model
E1). It is evident that the distribution is depleted of short merger
times. In contrast to the previous two models, the population is
dominated by BH-NS and BH-BH systems, which have in general wider
initial separations, and thus longer merger times than NS-NS binaries. 
Model H2 (no helium-star CE phases) shows a very similar behavior with the 
peak at short merger times disappearing. 

Other models lead to the double-peaked distributions similar to our 
standard-model results. The range of values remains unchanged, but the
relative strength of the two peaks vary slightly from model to model
following the variations of the dominant classes within the DCO
population.

\subsection{Results on Double Compact Object Coalescence Rates}

\subsubsection{Galactic Coalescence Rates of Double Compact Objects}

The calculated DCO coalescence rates have been calibrated using the
latest Type II SN empirical rates normalized to our Galaxy
\footnote{Following van der Kruit (1987) we have adopted a value of $2
\times 10^{10} {\rm L}_\odot$\ for the blue luminosity of the Galaxy, 
the value used in several other studies, e.g., Phinney (1991) or Portegies
Zwart \& Yungelson (1998). Some recent results point to a value lower by
a factor of $\simeq 2$ (see discussion in Kalogera et al.\ 2001, and
references therein). Had we adopted this lower value, all our
coalescence rates would also be decreased by the same factor of $\simeq
2$.} (Cappellaro, Evans, \& Turatto 1999).  In Table~\ref{rates1}
coalescence rates are given for all the models in our study for each of
the DCO classes as well as for the whole population. For our standard
model (A), rates were calculated from a very large number of primordial
binaries ($3\times 10^7$) and sub-samples of DCO formed from each of
$10^6$-binaries runs were used to examine and ensure the statistical
accuracy of the results (see \S\,3.2).  All other rates are based on
models with $N \geq 10^6$ primordial binaries. For each model, we have
also evolved an equal number of single stars, except for models K1-2,
for which we appropriately adjusted the contribution of single stars
based on the assumed binary fraction.

\begin{figure*}[t]
\centerline{ \psfig{file=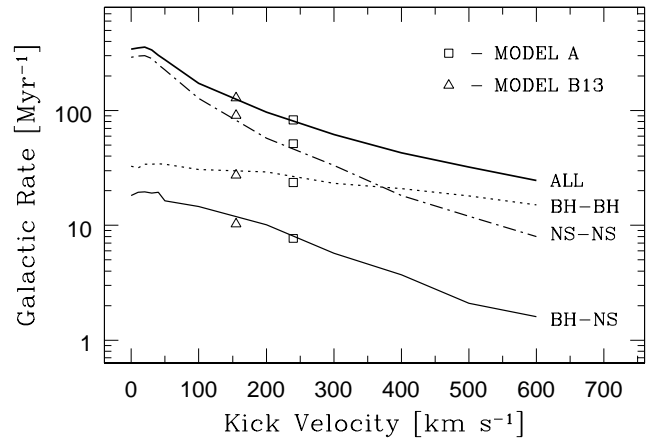,width=0.5\textwidth} }
\caption{
Dependence of Galactic coalescence rates on the assumed natal kick
velocity distribution. Lines connect rates for models B1-B12 and the
horizontal scale shows the width of Maxwellian kick distribution of a
given model. Points mark rates for our standard model (A) and model with
``Paczynski-like'' kick distribution (B13). The horizontal position for
these two models does not correspond to the horizontal axis scale, but 
instead is
chosen so that the predicted rates approximately match the curve for the
Maxwellian kick distributions.}
\label{fig10}
\end{figure*}

In Figure~\ref{fig10}, we illustrate the rate dependence on the assumed
kick velocity distribution. There is an overall decrease of rates with
increasing kick velocity for every population of coalescing double compact
objects, as the disruption probability for pre-collapse systems increases
with higher kick magnitudes.  It is worth noting though that at small
kicks (from zero to average magnitudes of $\sim 30$\,km\,s$^{-1}$) rates
are found to increase.  This is due to the importance of kicks in creating
tight binaries with merger times shorter than a Hubble time (see Fryer \&
Kalogera 1997 and Fryer et al.\ 1999 in the NS-NS case).  Rates remain 
roughly
constant for $\sigma = 20-50$\,km\,s$^{-1}$, as the two effects balance
one another. It is also notable that the rate decrease is steeper for
NS-NS binaries, and progressively flattens for BH-NS and BH-BH binaries.
This is a result of the lower kick magnitudes assumed to be imparted to
BH. For very high kick magnitudes (Maxwellians with $\sigma > 400 {\rm
km s}^{-1}$), BH-BH systems start dominating the population, as most of
the NS-NS progenitors get disrupted. These results are in qualitative
agreement with previous studies (e.g., Lipunov et al.\ 1997; Portegies
Zwart \& Yungelson 1998; Fryer et al.\ 1999).  In Figure~\ref{fig10}, we
also plot the rates from our standard model (kicks similar to the
results by Cordes \& Chernoff 1998) and the model B13 with
``Paczynski-like'' kicks. It is evident that, despite the very different
shape of these distributions, they closely correspond to Maxwellian
kicks with $\sigma \simeq 240 {\rm km\,s}^{-1}$ and $\sigma \simeq 150
{\rm km\,s}^{-1}$, respectively.  This result once again indicates that
the kick distributions are narrowly ``filtered'' by DCO binary
properties: only binaries receiving kicks similar to their orbital
velocities have a good SN survival chance (Kalogera 1996). The match
between these two models and the specific $\sigma$-values of the
Maxwellians indicates that the normalizations in the velocity range of
interest happen to be very similar and the shape of the distributions 
outside this range becomes irrelevant.

The effects of hyper-critical accretion combined with the assumed
maximum NS mass can be understood on the basis of the results of models
C and D1-2. We find that hyper-critical accretion in CE phases not only
increases the mass of the inspiraling compact objects and can convert NS
to BH, but also leads to somewhat wider post-CE systems because part of
the envelope is accreted and does not need to be expelled at the expense
of orbital energy. In model C, we do not allow for any hyper-critical
accretion and we find that the rate of NS-NS and BH-NS systems decreases
whereas the rate BH-BH binaries remains unaffected. This combination
indicates that it is the effect on the orbital period (increased rate of
post-CE mergers)  that dominates over the reduction of NS conversions to
BH. BH-BH systems originate from wider progenitors and therefore are not
much affected.  The reduction in NS conversions to BH turns out to be
unimportant because the assumed maximum NS mass is rather high and the
rate of conversions is low in the standard model (only 5\% of all NS
entering CE phases). In contrast, when the maximum NS mass is reduced to
2\,M$_\odot$ and 1.5\,M$_\odot$ (models D1 and D2), 34\% and 80\%
respectively of NS entering CE phases collapse into BH. In these two
models there is also a clear decrease in the rate of NS binaries and
increase of the BH-BH coalescence rates, whereas the total DCO
coalescence rates remain essentially constant (within our statistical
accuracy, see \S\,3.2).

The effects of varying the effective CE efficiency (models E1-3) are
qualitatively similar to not allowing for hyper-critical accretion
(model C).  Coalescence rates tend to decrease with decreasing CE
efficiency because of an increased rate of CE mergers.  The main
difference with model C is that the BH-BH rates are altered here.  This
is because varying the CE efficiency by factors of 2 or more affects
post-CE binary separations much more than decreasing envelope masses of
an ejected envelope by a few tenths of a solar mass in model C.

In models F1-2 and L1-2, we vary the parameters governing mass loss in
non-conservative MT episodes, and in particular in model F2, we obtain
results for conservative mass transfer.  Most of the NS-NS progenitor
systems ($\sim 75\%$) start their mass transfer history with a
non-conservative exchange, and therefore their coalescence rates might
be altered. Also for many other compact binary progenitors,
non-conservative mass transfer phases may take place, particularly at
the early stages of binary evolution, when the primary expands and
evolves toward the red giant branch while its companion is still on the
main sequence. However, non-conservative mass transfer phases do not
drastically change the orbital separation, so we find that the overall
coalescence rates do not change by much. Comparison to our standard
model shows a depletion in coalescing BH-BH systems for both F1-2 and
L1-2 models, and also moderate decrease of NS-NS binaries for the F1 and
L2 models. Coalescence rates of BH-NS binaries are not significantly
altered. The similarities of the results come from the same effect that
the fact that the F and L model parameters affect post-MT orbital
separation in a similar way. In models F, we vary $f_a$, the amount of
material which is accreted onto the companion star, while in models L,
we change the specific angular momentum $j$ of the material lost from
system during MT ($1.0 - f_a$). Both an increase of $f_a$ (F2) and
decrease of $j$ (L1) lead to an increase of the final post-MT orbital
separations. On the other hand, both a decrease of $f_a$ (F1) and
increase of $j$ (L2) lead to a decrease of the final post-MT
separations.  (see eq.~\ref{pod1} and eq.~\ref{pod2}, or original study
of Podsiadlowski et al.\ 1992). Note, that orbital separations also
depend on the mass ratios, but here we only discuss the changes relative
to the standard model. It turns out that these orbital separation
changes have different effects on different DCO populations. For NS-NS
progenitors, pre-MT orbits are already relatively tight, so in models
F1 and L2 (F2 and L1) the NS-NS rate decreases (increases) as the
frequency of mergers increases (decreases).  Progenitors of BH-BH
systems are more massive and generally have wider orbits at the onset of
MT episodes compared to the progenitors of NS-NS binaries. Given the
decreased post-MT separations of model F1 and L2, the BH-BH rate drops
as in the case of NS-NS systems and for the same reasons. However, for
the increased post-MT separations of model F2 and L1, the BH-BH rate
drops as well, this time due to the fact that many final BH-BH binaries
are not tight enough to coalesce within the Hubble time.

Variation of wind mass loss rates (models G1 and G2) affects more
strongly the rates of BH binaries, which decrease with stronger mass
loss. The extreme case of wind losses even stronger than in our standard
model entirely eliminates the BH-BH population, but increases the
overall DCO rate, mainly because survival through CE phases and SN
events is facilitated (lower-mass envelopes).

As we pointed out in \S\,2.1.2 (see also Belczynski \& Kalogera 2001),
the maximum mass of helium stars that develop convective envelope is
somewhat uncertain (thought to lie within the range of
$3.5-4.5$\,M$_\odot$), and therefore we treat it as a model parameter.
We find that reducing this value to $M_{\rm conv}=4.0 M_\odot$ (model H1)
results in NS-NS coalescence rate decreased by 30\%. This is expected
since many of the NS-NS formation channels (see Table~\ref{channels})
involve CE evolution of helium stars, which is aborted in the absence of
convective envelopes. If we further eliminate the possibility of CE 
evolution even for low-mass helium stars (model H2), the NS-NS 
coalescence rate drops to $0.9$\,Myr$^{-1}$. 

Although it is not thought to be relevant for our Galaxy, we examine one
model where star formation is assumed to be burst-like and to have
occurred 10\,Gyr ago (model I). We find that all coalescence rates are
increased because a larger fraction of them had enough time to evolve
and have total lifetimes shorter than 10\,Gyr.

As expected, a flatter IMF (model J) favors the formation of NS and even
more of BH, so the contribution of BH-BH and BH-NS systems increases as
the overall DCO coalescence rate increases as well.

Varying the binary fraction affects the overall normalization of our
population synthesis models. Binary systems mostly contribute to type
Ib/c SN and therefore an increased binary fraction decreases the
absolute number of Type II SN in the model and, for an assumed Type II
SN normalization (based on the empirical estimates), leads to an
increase of the coalescence rates.

In model N, we see that rates of systems containing BH are similar to
those of the standard model. However, the rates of NS-NS are much
depleted. This is due to the fact, that in model N, with no helium star
radial expansion, the NS-NS progenitors avoid the last CE episode of
standard model formation channels. Therefore, many NS-NS systems of
model N are not tight enough to merge within the Hubble time.

In models M, where we used different distributions for the initial binary
mass ratios, the total DCO coalescence rates are reduced or increased,
depending on whether extreme mass ratios (small $q$ values) are favored or
not.  In model M1 the MT events are generally dynamically unstable,
leading very often to mergers long before compact object systems form.

Evolution of model O influences only systems containing a BH, as in this
model we have allowed for BH formation only through partial fall back.
In the standard model, some BH were formed through a direct collapse of
a massive star, and no kick was imparted to such BH nor any material was
lost from the binaries. Extending the influence of BH formation through
partial fall back increases the overall probability that binaries are
disrupted.  Thus coalescence rates of BH-NS and BH-BH binaries in model
O are lower.

\subsubsection{Rate Increase Factors for Empirical Estimates of NS-NS
Coalescence}

Coalescence rate estimates based on the observed sample of close NS-NS
systems need to account for all possible observational biases acting
against their detection. Systems with two NS are discovered as binary
pulsars. The two coalescing systems that have been observed in the
Galactic Observed sample of NS-NS systems contain recycled NS with long
pulsar lifetimes ($\sim 10^9$\ yr) and long merger times ($> 10^8$\ yr),
and empirical estimates are obtained under the assumption that the
observed systems are representative of the Galactic population (Kalogera
et al.\ 2001).

As emphasized in our earlier work (Belczynski \& Kalogera 2001),
empirical coalescence rate estimates could be increased in light of the
new short-lived NS-NS sub-populations identified in studies of their
formation.  {\em Rate-increase factors} can be calculated to account for
the populations of (i) non-recycled NS-NS, and (ii) tight NS-NS binaries
with very short merger times formed through the new formation channels
identified in the present study. Because of their highly reduced
lifetimes (pulsar and merger), the detection efficiency for these NS-NS
sub-groups is practically diminished (drops by $\sim 3$ orders of
magnitude).

In Table~\ref{Corr}, we present Galactic coalescence rates of: non-recycled
NS-NS, tight NS-NS binaries with merger times shorter than 1 Myr, and 
the complete NS-NS population. In the last column of Table~\ref{Corr} we 
give the correction factors for empirical coalescence rates calculated 
from: 
 \begin{equation}
 { {\cal R} \over {\cal R} - ({\cal R}_{\rm nr}+{\cal R}_{\rm tight})
}\, ,
 \end{equation}
 where ${\cal R}$ is the rate of all coalescing NS-NS binaries, ${\cal
R}_{\rm nr}$ is the rate of coalescing non-recycled NS-NS (group I),
${\cal R}_{\rm tight}$ is the rate of systems (of group II and III) with
merger times smaller than $1$\ Myr. These correction factors imply an
increase of the NS-NS rate estimated empirically based on the observed
NS-NS with long pulsar and merger lifetimes.

It is evident from Table~\ref{Corr} that systems formed through the new
``helium-star'' channels (group I and II) contribute the most to the
correction factors. Our standard model prediction is that empirical
rates may be increased by a factor of 2.5 to account for short-lived
NS-NS pulsar systems.  Although the correction factors change
significantly for a few of the models, i.e., from 1.4 to 6.1 for models
with different CE efficiency (E1--3), they remain roughly constant at
$\sim 2.5$ for most of the models in our parameter study. Only for the
nonphysical model N, with no helium star radial evolution, the
correction factor is found as expected to be 1.0. Note that even for 
model H2 (no helium-star CE evolution) there is a small fraction of 
short-lived systems (primarily due to kicks instead of CE orbital 
shrinkage) and the correction factor is 1.1. In model N no
correction is needed, as in this model we do not form any NS-NS binaries
in groups I nor II, and both these groups are primarily responsible for
the increase factors (see also footnote in \S\,2.1.2).

For our correction-factor estimates we have conservatively assumed (i)
that a recycled NS is formed in {\em all} NS-NS binaries in which the
first-born NS had an opportunity to interact with its non-degenerate
companion, even though this may not be true (given our limited
understanding of recycling we cannot be certain). We characterize
non-recycled NS-NS only systems that evolved through a double CE phase
involving two helium stars (channels NSNS:09, NSNS:11, and NSNS:12);  
(ii) that systems with merger times longer than 1\,Myr can detected with
the same efficiency as much longer lived binaries, even though in
reality there is a continuum, given the continuous distribution of
merger times (Figure~\ref{fig7}). Because of these two assumptions we
should consider the derived upwards correction factors to be rather
conservative and represent more of lower limits than rough estimates.

\subsubsection{Predicted Supernova and Star Formation Rates}

Cappellaro et al. (1999) estimated the rates of Type II SN and Type Ib/c
SN to be $0.86 \pm 0.35$ SNu and $0.14 \pm 0.07$ SNu for Sbc-d galaxies,
where 1 SNu corresponds to one SN per 100 yr and the estimates are
normalized to a blue luminosity of $10^{10}{L_\odot}^B$.  For an
estimated Galactic blue luminosity of about ${L_\odot}^B=2 \times
10^{10} L_\odot$ (van der Kruit 1987; although see footnote in
\S\,3.5.1), we obtain 1.72 and 0.28 SNu for Type II and Ib/c SN,
respectively.  The ratio of Type II to Type Ib/c SN turns out to be
quite uncertain: $6.1 \pm 4.0$. Note that this ratio is independent of
the assumed blue luminosity for our Galaxy.

In our population synthesis calculations we keep track of all SN events.
We use Type II events to normalize our models and then combine them with
the Type Ib/c rates to derive the model-predicted II-to-Ib/c ratios. We
find them to lie in the range $1.8-3.7$, for all the models in our
parameter study, with the standard model yielding a ratio of 2.6.
Although our predicted ratio values are consistent with the empirical
estimates within the associated errors, we note that the theoretical
values tend to be systematically lower than the empirical estimates,
implying that our Ib/c rates are rather high. In retrospect, this is
actually expected given some of the assumptions in the MT events. We
note that in dynamically stable and unstable MT episodes, we assume that
the donors are always stripped of their envelopes. However, it is not at
all clear that this is a realistic assumption in small cases and it is
quite possible that a small fraction of the H-rich envelope remains with
the post-MT/CE donors. Since our classification of SN events as type
Ib/c is based on the exploding stars having lost their H-rich and/or
He-rich envelopes, it is possible that we are overestimating the Ib/c
events because of this assumption. We note, however, that all our models
remain consistent with the empirical rates within the estimated errors. 

Another explanation for the possible overabundance of Type Ib/c SN in
population synthesis model has been provided by De Donder \& Vanbeveren
(1998).  Our results are consistent with their findings of II-to-Ib/c
ratios of 2-3. The explanation they offered was related to possible
variations of massive binary formation in galaxies over a range of
morphological types and to the empirical estimates representing an
average over many galaxies and not being appropriate for our Galaxy.
However, in their study De Donder \& Vanbeveren (1998) considered the
results obtained by (Cappellaro et al. 1993a,b) that were available at
the time. In contrast, the most recent study of empirical SN rates
(Cappellaro et al. 1999) accounts for different morphological types of
galaxies, possibly implying that the MT assumptions (common to our and
the De Donder \& Vanbeveren 1998 study) may be primarily responsible for
the calculated SN small rate ratios.

Galactic SFR have been estimated to lie in somewhat broad ranges of $1-3
{\rm M}_\odot {\rm yr}^{-1}$ (Blitz 1997; Lacey \& Fall 1985) and $\sim
1-10 {\rm M}_\odot {\rm yr}^{-1}$ (Gilmore 2001).  We use the
calibration to the Type II SN rate by Cappellaro et al. (1999)  with the
adopted blue luminosity of our Galaxy: $2 \times 10^{10} {\rm L}_\odot$
(van der Kruit 1987) to calculate the Galactic SFR corresponding to our
models. To do so we have to make an assumption about the extension of
the IMF down to 0.08\,M$_\odot$. If we assume that the IMF continues as
a steep power-law down to the hydrogen-burning mass limit, we obtain
values in the range $7-28 {\rm M}_\odot {\rm yr}^{-1}$, which is clearly
significantly higher than current estimates for the Galactic SFR.  
Based on the results of Scalo (1986, see also Kroupa, Tout, \& Gilmore
1993), the IMF is thought to flatten for masses below about
1.0\,M$_{\odot}$. With the assumed flattened IMF of Kroupa et al.\
(1993), we obtain much lower SFR values in the range of $3-9 {\rm
M}_\odot {\rm yr}^{-1}$, with the standard model prediction of $\sim 6
{\rm M}_\odot {\rm yr}^{-1}$.  Moreover, had we adopted the lower, by a
factor of two, blue luminosity of our Galaxy (see Kalogera et al.\ 2001,
and references therein) the predicted SFR would decrease (by the factor
of two) even further. We note the good agreement of the Galactic SFR
estimates with our predictions for a flattened IMF.

\subsubsection{Comparison With Other Studies}

In \S\,3.4 we compared our results for the physical properties of DCO
populations to those of earlier studies wherever possible. In this
subsection we explicitly focus on such a comparison based on results for
coalescence rates from the following studies:  Lipunov et al.\ (1997), 
Portegies Zwart \& Yungelson (1998, hereafter PZY), De
Donder \& Vanbeveren (1998, hereafter DDV), Fryer et al. (1999,
hereafter FWH), and Nelemans et al.\ 2001b.

Lipunov et al.\ (1997) focused their study on the effect of SN kicks on
coalescence rates of double compact objects. Their calculations showed
that, in general, the rates decrease approximately exponentially with
increasing kick magnitude. This finding is in very good agreement with
our calculations (see \S\,3.5.1 and Figure~\ref{rates1}). We also see
that the behavior of {\em relative} coalescence rates of NS-NS to BH-NS
binaries is quite similar to what we obtain here. In both studies, the
ratio of rates (NS-NS to BH-NS) decreases with increasing average kick
velocity.  However, the absolute coalescence rates differ and this is
easily understood as a result of different population synthesis
assumptions, important for the calculation of coalescence rates. In
particular, Lipunov et al.\ (1997) use quite different compact object
formation scenarios, as well as a different IMF.

PZY studied the formation and evolution of NS-NS and BH-NS binaries and
included a limited parameter study. Nelemans et al.\ (2001b) calculated
populations of Galactic binaries with white dwarfs, NS, and BH, and
estimated the low-frequency gravitational-wave emission of these
binaries. They used a population synthesis code similar to that of PZY,
with few modifications concerning BH formation. Although, many of the
population synthesis assumptions differ, we find some similarities in
the results. In general, rates of NS-NS binaries are higher than those
of the BH-NS systems. However, in all PZY models NS-NS dominate over
BH-NS binaries, which is not true for all of our models mainly because
of our more extensive parameter study.  Moreover, on average the rates
obtained by PZY are smaller than ours, but for some comparable models
they differ only by factors of a few.  This is due to the newly
identified population of NS-NS binaries, as well as to the other
differences in model assumptions. PZY note that due to hyper-critical
accretion onto NS in CE phases, the population of BH-NS may dominate
over NS-NS systems. Although, they do not perform actual calculations of
hyper-critical accretion events (they assume that NS always collapse to
BH), their results are in agreement with ours, provided that the maximum
NS mass is rather low (smaller than 2.0 $M_\odot$, see rates of models
D1 and D2 in Table~\ref{rates1}). The predicted BH-BH coalescence rates
are diminishingly small in both studies (BH-BH binaries form but in very
wide orbits with merger times longer than a Hubble time). This is the
combined result of a number of factors: in these two studies BH are
assumed to form (i)  symmetrically and no birth kicks are taken into
account, and (ii)  only from progenitor stars more massive than
40\,M$_\odot$, which biases the population to lower frequencies {\em
and} wider orbits.

With respect to the DDV study, we note the difference between formation
rates that correspond to the entire population and coalescence rates
that correspond to the subgroup with coalescence times shorter than a
Hubble time. DDV reported formation rates and hence a direct comparison
with their results is difficult. Overall their formation rates are very
high (not only compared to our results but also all the other population
studies): $\sim 400$\ NS-NS, $\sim 2000$\ BH-NS and $\sim 140000$\ BH-BH
systems per Myr in Galaxy. DDV comment that their BH-BH formation rate
is surprisingly high. One reason may be their calibration method based
on an assumed massive star formation rate of one massive star per year.
This does not easily transform into a star formation rate as used in
this work and consequently makes the comparison very difficult.

FHW focused their work to the study of gamma-ray burst progenitors and 
among them NS-NS and BH-NS binaries. 
They performed quite an extensive parameter
study and calculated coalescence rates for all their models. Once again,
we note the very similar dependence of their rates and ours on the
average kick magnitude.  One striking difference comes with the very
wide ranges of their predicted rates (e.g., their 4 orders of magnitude
change of NS-NS coalescence rate compared to about 2 orders of magnitude
found here).  The enormous spread of rates in the FHW work comes from
the fact, that in one of their models they, non-physically, increased
(by factor of four) the maximum stellar radii to explore the uncertainty
related to radius determination in evolved stars. Such a change
significantly affects the evolution of many binaries, since the stellar
radius is a crucial quantity in judging on the occurrence or
non-occurrence of one of the most important binary processes: MT events.
The radii we calculate for stars come from the stellar evolution models,
and they depend on the star mass, its evolutionary stage, and its
composition, and they are not altered in any nonphysical way. However,
the FWH models of varied radii show the sensitivity of the results of
population synthesis to this parameter, which is not very well
constrained by stellar evolution models.

The results on coalescing compact object binary rates are in good
qualitative agreement with previous theoretical predictions. Although we
have noted some systematic differences we are able to attribute them to
the different population synthesis model assumptions, as well as to the
effects of the newly identified populations of NS-NS binaries.

\section{DISCUSSION}

Close binaries of NS and BH have attracted an increased interest in
recent years primarily because of their connection to gravitational-wave
detection and possibly to gamma-ray burst progenitors. Here we focus the 
discussion of our results in the context of gravitational-wave detection
by the upcoming ground-based interferometers (e.g., LIGO) and the
prospects of detecting inspiraling DCO. An analysis of these and other
populations in connection to gamma-ray bursts is presented in 
Belczynski, Bulik \& Rudak 2002c and Perna \& Belczynski 2002.

Our motivation in initiating this study was to examine DCO populations
in view of some developments in the understanding of CE evolution,
particularly the possibility of double CE ejection (suggested by Brown
1995) and hypercritical accretion (suggested and quantified by Chevalier
1989, 1993; Brown 1995; Bethe \& Brown 1998, and in the present study),
and the evolution of low-mass helium stars. Some, although not all, of
these possibilities have been explored in some of the earlier population
synthesis studies, but not in a consistent and equally detailed way
(mostly some extreme cases have been examined; for example, either
hyper-critical accretion has been ignored or all NS entering a CE have
been assumed to collapse to BH). Our goal is primarily two-fold: (i) to
examine the predicted rates for various DCO classes, focusing mainly on
the relative formation frequencies and their behavior with a large
number of model parameters, and (ii) to examine in detail the physical
properties of the various populations, the origin of their
characteristics, and the links to certain key evolutionary effects or
phases, and to identify the most robust of the qualitative features in
the distributions of binary properties.

In the course of our investigation we came across a number of new
formation channels leading to rather efficient formation of NS-NS
binaries. Apart from an increase in the predicted rates, examination of
their properties revealed that in their majority these NS-NS binaries
form a very distinctive class with tight orbits and short merger
timescales (i.e., lifetimes).  The common thread connecting all these
new formation paths is the evolution of low-mass helium stars and the
implications of the fact that they can develop partially or fully
convective envelopes.  We find that a large number of MT episodes
initiated by these evolved stars lead to high survival rate through a CE
phase and the production of very tight binaries, which in turn have very
high survival probability through the second core-collapse event.  
Further, following the basic argument made by Brown (1995) we have
allowed for the possibility of two giant stars (He-rich more relevant to
efficient DCO formation, but also H-rich) involved in the dynamically
unstable MT to eject the two combined common envelopes.  It is these
possibilities that lead to qualitative changes in the DCO population
characteristics, which have been discussed above.

It is important to acknowledge that the viability of these new formation
channels has not been examined with detailed evolutionary and
hydrodynamical calculations. Despite the difficulties of fully
understanding the details of CE phases, questions such as: ``are DCE
events realistic?'', ``can evolved helium stars survive a CE phase?'', and
``under what conditions?'', are currently traceable (model H2 corresponds
to the case of the answers to the above questions being no for helium
stars). It seems that such investigations are necessary before we can
include these new channels as part of what is considered ``standard'' ways
of forming DCO.  Nevertheless it is interesting to examine the
implications of such evolutionary phases (see also Fryer et al.\ 1999;
Nelemans et al.\ 2001a).

In the context of gravitational-wave detection from inspiraling NS-NS
binaries, these relatively short-lived systems imply that empirical
rates derived based on the observed sample (with much longer lifetimes)
should possibly be raised by factors of 2-3 typically. The reason is
that they correspond to a Galactic NS-NS population that is not
represented in the observed binary pulsar sample but would very well
contribute to inspiral events.  Using the results of Kalogera et al.\
(2001) for the empirical NS-NS coalescence rate, we find that their {\em
most optimistic} prediction for the LIGO I detection rate could be
raised to at least 1 event per 2 years, and their {\em most pessimistic}
LIGO II detection rate could be raised to 5 events per year or even
higher.

In Table~\ref{ligo1} and Table~\ref{ligo2} we present our theoretical
predictions of detection rates of different binary merger types for LIGO
I and LIGO II, respectively.  These detection rates correspond to our
coalescence rates of double compact objects calculated with the {\em
StarTrack} population synthesis program. Using the extragalactic
extrapolation of Kalogera et al.\ (2001) and the maximum sensitivity
distances of LIGO I and LIGO II for a given binary merger type (see
Kalogera \& Belczynski 2001), we converted our Galactic coalescence
rates to detection rates. We find that the LIGO I detection rates are
quite low with a {\em maximum} total rate of a couple events per year
and with rates significantly below 1 event per year for many models.  
However, we find that the prospects for double compact object inspiral
detection are very encouraging for LIGO II, for which we predict at
least $\simeq 10$ detections per year, even in the most pessimistic
case. Moreover, for LIGO II, the total detection rate of compact object
binaries is as high as few hundred events per year for most of our
models.

The uncertainties of population synthesis method (reflected on the
predicted detection ranges) seem much reduced compared to the earlier
results (see Fryer et al.\ 1999; Kalogera et al.\ 2001).  This is
related to the fact that in the present study we adopted a given set of
single star evolution models (however note that we did explore the
effects of varying wind mass-loss rates and helium-star evolution), and
we have adopted a physically motivated initial-remnant mass relation
based on hydrodynamical calculations of core-collapse events. In light
of some indicative results from rotating star models (Heger, Langer \&
Woosley 2000; Heger \& Langer 2000), we can expect stellar evolution
models to be updated in the future and we would regard this as progress
in the field.

We regard the investigation of the physical properties of DCO as
important as that of the rates and quite revealing in terms of their
origin and robustness.  The details have already been discussed earlier
in the paper. From the point of view of gravitational-wave astrophysics,
we hope these results will open new directions in anticipating and
understanding properties of gravitational-wave sources.  For example,
distributions over components masses can be used to produce fake
inspiral data to test the data analysis tools currently under
development or evaluate detection efficiencies (we are currently
involved in such an activity already planned within the LIGO Scientific
Collaboration as part of the preparation for the LIGO I Scientific run
in 2002). Predictions for physical properties can be used in developing
specialized data-analysis tools to explore an astrophysically motivated
parameter space, in cases where the unrestricted parameter space of
inspiral signals is just far beyond any current or near-future
computational capabilities (see Kalogera 2000). Looking a little more
into the future, in the era when gravitational-wave astronomy is
possible and physical properties are measurable, the identification of
strong qualitative features (such as those discussed here: relative
contributions of DCO classes, or presence or absence of certain peaks
and other features) will allow us to evaluate some of the surviving
theoretical models.

\acknowledgements 
 We thank Jarrod Hurley for many very useful comments on single star
evolution, Teviet Creighton for discussions of LIGO project needs,
Jeffrey McClintock for comments on the observed compact object mass
distribution, Chris Fryer for his population synthesis and
common-envelope insights, Ron Taam for his help in our common-envelope
understanding, and Thomas Matheson for discussion of SN rates. KB
acknowledges support from the Smithsonian Institution through a
Predoctoral Fellowship, from the Lindheimer fund at Northwestern
University, from the Polish Science Foundation (FNP) through a 2001
Polish Young Scientist Award, and from the Polish Nat.\ Res.\ Comm.\
(KBN) through grants 2P03D02219 and 5P03D01120. VK acknowledges support
from the Smithsonian Institution through a Clay Fellowship and from NSF
Grant PHY-0121420. TB acknowledges support from the Polish Nat.\ Res.\
Comm.\ (KBN) through grant 5P03D01120. VK and TB are also grateful for
the hospitality and support of the Aspen Center for Physics.

\newpage

\begin{appendix}

\section{HYPER-CRITICAL ACCRETION ONTO COMPACT OBJECTS DURING 
COMMON ENVELOPE PHASE}

Let us denote the mass of the compact object (NS or BH) by $M_{\rm A}$,
the mass of its companion (H-rich or He-rich giant) by $M_{\rm B}$, its
core mass by $M_{\rm B,core}$, its radius by $R_{\rm B,ce}$, and the
binary semi-major axis by $A_{\rm ce}$. Accretion onto the compact
object can be initiated only once the binary separation becomes equal to
the radius of the expanding giant donor. Both quantities have changed
since the time of Roche-lobe filling to $A_{\rm acc}$ and $R_{\rm
B,acc}$, so that $A_{\rm acc}=R_{\rm B,acc}$. CE evolution and accretion
onto the compact object will end when the giant's envelope has been
ejected and the donor's mass changes from $M_{\rm B,i}$ to $M_{\rm
B,f}=M_{\rm B,core}$. Throughout the derivation $\alpha_{\rm ce}$ is the
CE efficiency parameter and $\lambda$ is the numerical factor scaling
the binding energy of the donor. The orbit is expected to be circular
due to circularization on very short time scale as the donor approached
Roche-lobe filling.

Following Bethe \& Brown (1998) we write the energy loss rate related 
to the accretion onto the compact object
 \begin{equation}
 \dot{E}_{\rm acc} = 0.5 c_{\rm d} G (M_{\rm B}+M_{\rm A}) A^{-1} 
\dot{M}_{\rm A}
 \end{equation}
 where $c_{\rm d}$ is the drag coefficient of the compact object with
respect to the donor's envelope, and it is assumed $c_{\rm d}=6$ (Shima
et al. 1985). The rate of orbital energy dissipation is given by:
 \begin{equation}
 - \dot{E}_{\rm orb} = 0.5 G M_{\rm B} A^{-1} \dot{M}_{\rm A} + 
                      0.5 G M_{\rm A} A^{-1} \dot{M}_{\rm B} - 
                      0.5 G M_{\rm A} M_{\rm B} A^{-2} \dot{A}                         
 \end{equation}

Orbital energy is dissipated due to the dynamical friction of accreting 
neutron star in the giant envelope.
Thus, we may compare equations A.1 and A.2 ($\dot{E}_{\rm acc} = - 
\dot{E}_{\rm orb}$) 
and after taking derivative in respect to $dM_{\rm B}$ we obtain,
 \begin{equation}
 [c_{\rm d} (M_{\rm B}+M_{\rm A}) - M_{\rm B}] {dM_{\rm A} \over dM_{\rm 
B}} = - M_{\rm A} M_{\rm B} A^{-1} {dA \over dM_{\rm B}} + M_{\rm A}
 \end{equation} 
During the phase of accretion we may express binding energy of donor 
envelope as,
 \begin{equation}
 - E_{\rm bind} = G M_{\rm B} (M_{\rm B} - M_{\rm B,core}) \lambda^{-1} 
                 A^{-1} 
 \end{equation}
 where instead of the donor radius we use the binary separation, which
at the start of the accretion phase is equal to the donor radius
($A_{\rm acc}=R_{\rm B,acc}$).  
The rate of binding energy change can be written as,
 \begin{equation}                                                        
 - \dot{E}_{\rm bind} = G \lambda^{-1} A^{-1} (2 M_{\rm B} - M_{\rm
B,core}) \dot{M}_{\rm B} + G M_{\rm B} \lambda^{-1} A^{-2} (M_{\rm
B,core} - M_{\rm B}) \dot{A}
 \end{equation}

The donor envelope is ejected on the expense of the binary orbital 
energy, with an efficiency described by parameter $\alpha_{\rm ce}$.
Thus, we write CE energy balance with respect to the donor mass as,
 \begin{equation}
 - \alpha_{\rm ce} (- {dE_{\rm orb} \over dM_{\rm B}}) = 
(- {dE_{\rm bind} \over dM_{\rm B}})
 \end{equation}
 We use equations A.2, A.5, and A.6 to write:  
 \begin{equation}
 M_{\rm B} {dM_{\rm A} \over dM_{\rm B}} = M_{\rm B} A^{-1} [- 2 
\lambda^{-1}
\alpha_{\rm ce}^{-1} (M_{\rm B,core} - M_{\rm B}) + M_{\rm A}] {dA \over dM_{\rm B}} 
- 2 \lambda^{-1} \alpha_{\rm ce}^{-1} (2 M_{\rm B} - M_{\rm B,core}) -
M_{\rm A} 
 \end{equation}
 We can next write out two ordinary differential equations to be solved: 
 \begin{equation}
 {dA \over dM_{\rm B}} = { M_{\rm A} M_{\rm B} h_1^{-1} + 2 \lambda^{-1}
\alpha_{\rm ce}^{-1} (2 M_{\rm B} - M_{\rm B,core}) + M_{\rm A} \over 
M_{\rm A} M_{\rm B}^2 A^{-1} h_1^{-1} + M_{\rm B} A^{-1} [- 2 \lambda^{-1}
\alpha_{\rm ce}^{-1} (M_{\rm B,core} - M_{\rm B}) + M_{\rm A}]
}
 \end{equation}
 \begin{equation}
 {dM_{\rm A} \over dM_{\rm B}} = 
M_{\rm A} h_1^{-1} (1 - M_{\rm B} A^{-1} f_1(M_{\rm A},M_{\rm B},A))
 \end{equation} 
 where $h_1=c_{\rm d} (M_{\rm B} + M_{\rm A}) - M_{\rm B}$, and
$f_1(M_{\rm A},M_{\rm B},A)$ is equal to the right hand side of equation
(A.8). We know the initial and final values for $M_{\rm B}$, the 
initial value for $M_{\rm A}$, and we can calculate the initial value 
for $A$ when accretion is initiated. To do so we use 
the CE energy balance for the pre-accretion period 
($- \alpha_{\rm ce} \Delta E_{\rm orb} = \Delta E_{\rm bind}$)
and obtain: 
 \begin{equation}
 A_{\rm acc}= {2 (M_{\rm B} - M_{\rm B,core}) + \lambda \alpha_{\rm ce} 
M_{\rm A}  \over 2 (M_{\rm B} - M_{\rm B,core}) R_{\rm B}^{-1} + \lambda
\alpha_{\rm ce} M_{\rm A} A_{\rm ce}^{-1}}
 \end{equation}
 where we set $R_{\rm B,acc}=A_{\rm acc}$. Equations A.8 and A.9
describe the accretion phase during CE evolution ($A<A_{\rm acc}$), and
we integrate them numerically from $M_{\rm B,i}$ to $M_{\rm B,core}$ to
obtain the final binary separation and final mass of the accreting
compact object.

\end{appendix}

\pagebreak

\begin{deluxetable}{ll}
\tablewidth{350pt}
\tablecaption{Population Synthesis Model Assumptions}
\tablehead{ Model & Description}
\startdata
A      & standard model described in \S\,2.1 and \S\,2.2 \\
B1--13 & zero kicks, single Maxwellian with \\
       & $\sigma=10,20,30,40,50,100,200,300,400,500,600$\,km\,s$^{-1}$, \\
       & ``Paczynski'' kicks with $\sigma=600$\,km\,s$^{-1}$ \\
C      & no hyper-critical accretion onto NS/BH in CEs \\
D1--2  & maximum NS mass: $M_{\rm max,NS}=2, 1.5$\,M$_\odot$ \\
E1--3  & $\alpha_{\rm CE}\times\lambda = 0.1, 0.5, 2$ \\   
F1--2  & mass fraction accreted: f$_{\rm a}=0.1, 1$ \\
G1--2  & wind changed by\ $f_{\rm wind}=0.5, 2$ \\
H1--2  & Convective Helium giants: $M_{\rm conv}=4.0, 0$$\,M_\odot$ \\ 
I      & burst-like star formation history \\
J      & primary mass: $\propto M_1^{-2.35}$ \\
K1--2  & binary fraction: $f_{\rm bi}=0.25, 075$ \\ 
L1--2  & angular momentum of material lost in MT: $j=0.5, 2.0$\\
M1--2  & initial mass ratio distribution: $\Phi(q) \propto q^{-2.7}, q^{3}$\\
N      & no helium giant radial evolution\\
O      & partial fall back for $5.0 < M_{\rm CO} < 14.0 \,M_\odot$\\ 
\enddata
\label{models}
\end{deluxetable}

\begin{deluxetable}{cccc}
\tablewidth{350pt}
\tablecaption{Population Synthesis Accuracy} 
\tablehead{ Coalescing & Mean & Max Change & Mean \\
           Population & Rate\tablenotemark{a} & in Rate\tablenotemark{a} & Number}
\startdata
NS-NS& 52.7 \ $\pm$ 1.1 (2\%) & 4.4 \hspace*{0.05cm} (8\%) & 1754 \\
BH-NS& 8.1  \ $\pm$ 0.6 (7\%) & 2.2 \hspace*{0.05cm} (27\%) &  269 \\
BH-BH& 25.6 \ $\pm$ 0.8 (3\%) & 3.9 \hspace*{0.05cm} (15\%) &  852 \\
Total& 86.3 \ $\pm$ 1.2 (1\%) & 5.5 \hspace*{0.05cm} (6\%) & 2875 \\
\enddata
\label{stat}
\tablenotetext{a}{Galactic Coalescence Rate (Myr$^{-1}$).}
\end{deluxetable}

\begin{deluxetable}{ccc}
\tablewidth{380pt}
\tablecaption{Double Compact Object Formation Channels - Standard Model}
\tablehead{ Formation & Relative   &  \\
            Channel & Efficiency \tablenotemark{a} & Evolutionary History
\tablenotemark{b}  }
\startdata

NSNS:01& 20.3\ \%& 
NC:a$\rightarrow$b, SN:a, HCE:b$\rightarrow$a, HCE:b$\rightarrow$a, SN:b\\

NSNS:02& 10.8\ \%&
NC:a$\rightarrow$b, SCE:b$\rightarrow$a, NC:a$\rightarrow$b, SN:a, 
HCE:b$\rightarrow$a, SN:b\\

NSNS:03& 5.5\ \%& 
SCE:a$\rightarrow$b, SN:a, HCE:b$\rightarrow$a, HCE:b$\rightarrow$a, SN:b\\

NSNS:04& 4.0\ \%&
NC:a$\rightarrow$b, SCE:b$\rightarrow$a, SCE:b$\rightarrow$a, SN:b,
HCE:a$\rightarrow$b, SN:a\\

NSNS:05& 3.2\ \%&
DCE:a$\rightarrow$b, SCE:a$\rightarrow$b, SN:a, HCE:b$\rightarrow$a, SN:b\\

NSNS:06& 2.5\ \%&
SCE:a$\rightarrow$b, SCE:b$\rightarrow$a, NC:a$\rightarrow$b, SN:a,
HCE:b$\rightarrow$a, SN:b\\

NSNS:07& 2.2\ \%&
NC:a$\rightarrow$b, NC:a$\rightarrow$b, SN:a, HCE:b$\rightarrow$a,
HCE:b$\rightarrow$a, SN:b\\

NSNS:08& 2.0\ \%& 
NC:a$\rightarrow$b, DCE:b$\rightarrow$a, SN:a, HCE:b$\rightarrow$a, SN:b\\

NSNS:09& 2.0\ \%&
DCE:a$\rightarrow$b, DCE:a$\rightarrow$b, SN:a, SN:b\\

NSNS:10& 1.6\ \%&
NC:a$\rightarrow$b, SCE:b$\rightarrow$a, SN:b, HCE:a$\rightarrow$b, SN:a\\

NSNS:11& 1.5\ \%&
NC:a$\rightarrow$b, SCE:b$\rightarrow$a, DCE:b$\rightarrow$a, SN:a, SN:b\\

NSNS:12& 1.5\ \%&
NC:a$\rightarrow$b, SCE:b$\rightarrow$a, DCE:a$\rightarrow$b, SN:a, SN:b\\

NSNS:13& 1.0\ \%&
DCE:a$\rightarrow$b, SN:a, HCE:b$\rightarrow$a, SN:b\\

NSNS:14& 3.0\ \%&
all other \\

&&\\

BHNS:01& 4.5\ \%&
NC:a$\rightarrow$b, SN:a, HCE:b$\rightarrow$a, SN:b\\

BHNS:02& 1.6\ \%&
NC:a$\rightarrow$b, SCE:b$\rightarrow$a, SN:a, SN:b\\

BHNS:03& 1.3\ \%&
SCE:a$\rightarrow$b, SN:a, HCE:b$\rightarrow$a, NC:b$\rightarrow$a, SN:b\\

BHNS:04& 2.0\ \%&
all other\\

&&\\

BHBH:01& 17.7\ \%&
NC:a$\rightarrow$b, SN:a, HCE:b$\rightarrow$a, SN:b\\

BHBH:02& 10.5\ \%&
NC:a$\rightarrow$b, SCE:b$\rightarrow$a, SN:a, SN:b\\

BHBH:03& 1.4\ \%&
all other \\

\enddata
\label{channels}
\tablenotetext{a}{Normalized to the total DCO population.}
\tablenotetext{b}{Sequences of different evolutionary phases for the primary
(a) and the secondary (b):
non-conservative MT (NC), single common envelope (SCE), double common
envelope (DCE), common envelope with hyper-critical accretion (HCE),
supernova explosion/core-collapse event (SN).
Arrows mark direction of MT episodes.}
\end{deluxetable}

\begin{deluxetable}{crrrr}
\tablewidth{300pt}
\tablecaption{ Galactic Double Compact Object Coalescence Rates
(Myr$^{-1}$)}
\tablehead{ Model\tablenotemark{a}& NS-NS& BH-NS& BH-BH& Total } 
\startdata

A   & 52.7  & 8.1  & 25.6 & 86.3  \\
B1  & 292.4 & 18.2 & 32.7 & 343.2 \\
B2  & 299.6 & 19.4 & 31.8 & 350.8 \\
B3  & 302.2 & 19.6 & 34.2 & 356.0 \\
B4  & 285.2 & 19.1 & 34.2 & 338.5 \\
B5  & 251.0 & 19.5 & 34.3 & 304.7 \\
B6  & 226.8 & 16.4 & 34.1 & 277.3 \\
B7  & 128.1 & 14.6 & 30.7 & 173.3 \\
B8  & 57.5  & 10.1 & 29.2 & 96.8  \\
B9  & 33.2  & 5.7  & 23.2 & 62.1  \\
B10 & 18.2  & 3.7  & 21.0 & 42.9  \\
B11 & 12.0  & 2.1  & 18.1 & 32.2  \\
B12 & 8.0   & 1.6  & 15.1 & 24.6  \\
B13 & 91.0  & 10.3 & 27.3 & 128.6 \\
C   & 43.2  & 5.6  & 23.2 & 72.1  \\
D1  & 33.6  & 23.3 & 31.1 & 88.0  \\
D2  & 9.1   & 36.2 & 42.2 & 87.5  \\
E1  & 2.7   & 4.8  & 5.6  & 13.1  \\
E2  & 23.5  & 6.3  & 23.1 & 53.0  \\
E3  & 109.0 & 8.7  & 11.5 & 129.2 \\
F1  & 22.1  & 9.3  & 8.7  & 40.1  \\
F2  & 54.3  & 8.6  & 7.2  & 70.1  \\
G1  & 43.9  & 14.2 & 75.6 & 133.7 \\   
G2  & 92.2  & 1.3  & 0.0  & 93.5  \\
H1  & 37.9  & 7.8  & 26.6 & 72.3  \\
H2  &  0.9  & 6.0  & 26.3 & 33.2  \\
I   & 54.5  & 10.0 & 33.6 & 98.1  \\
J   & 58.1  & 12.8 & 41.9 & 112.8 \\
K1  & 22.5  & 3.4  & 10.4 & 36.2  \\
K2  & 90.2  & 13.5 & 41.6 & 145.4 \\
L1  & 78.9  & 9.2  & 10.0 & 98.1  \\
L2  & 12.0  & 6.2  & 10.5 & 28.7  \\
M1  & 6.2   & 4.0  & 5.8  & 16.0  \\
M2  & 114.2 & 8.4  & 31.5 & 154.1 \\
N   & 34.4  & 10.7 & 24.5 & 69.6  \\
O   & 51.9  & 5.7  & 4.0  & 61.6  \\

\enddata
\label{rates1}
\tablenotetext{a}{for definition of models see Table~\ref{models}}
\end{deluxetable}

\begin{deluxetable}{ccrcrc}
\tablewidth{400pt}
\tablecaption{NS-NS Empirical Coalescence Rate Correction Factors}
\tablehead{  & Group I& Group II& Group III& Total& Empirical Rate \\
 Model\tablenotemark{a} & all& $t_{\rm merg}<1$\ Myr& $t_{\rm merg}<1$\ Myr& Rate&
Correction Factor} 
\startdata

A   & 4.4&  26.9& 0.1&  52.7& 2.5 \\
B1  & 6.0& 146.7& 0.0& 292.4& 2.1 \\
B6  & 6.5& 121.6& 0.0& 226.8& 2.3 \\
B7  & 5.2&  68.0& 0.1& 128.1& 2.3 \\
B8  & 4.7&  27.6& 0.2&  57.5& 2.3 \\
B9  & 4.3&  17.0& 0.1&  33.2& 2.8 \\
B10 & 2.8&   8.8& 0.1&  18.2& 2.8 \\
B11 & 2.2&   6.1& 0.0&  12.0& 3.3 \\
B12 & 1.9&   4.2& 0.0&   8.0& 4.2 \\
B13 & 4.6&  48.4& 0.1&  91.0& 2.4 \\
C   & 3.2&  22.1& 0.2&  43.2& 2.4 \\
D1  & 4.9&  15.1& 0.0&  33.6& 2.5 \\
D2  & 3.6&   2.8& 0.0&   9.1& 3.3 \\
E1  & 0.4&   1.9& 0.0&   2.5& 6.1 \\
E2  & 3.1&  15.1& 0.1&  23.5& 4.5 \\
E3  & 5.2&  28.0& 0.2& 109.0& 1.4 \\
F1  & 2.3&  12.2& 0.1&  22.1& 2.9 \\
F2  & 2.3&  16.1& 0.8&  54.3& 1.5 \\
G1  & 3.3&  23.7& 0.1&  43.9& 2.6 \\
G2  & 7.3&  38.0& 0.2&  92.2& 2.0 \\
H1  & 3.3&  20.9& 0.0&  37.9& 2.8 \\
H2  & 0.0&   0.0& 0.1&   0.9& 1.1 \\
I   & 4.0&  28.5& 0.0&  54.5& 2.5 \\   
J   & 4.4&  29.4& 0.1&  58.1& 2.4 \\
K1  & 1.8&  11.2& 0.1&  22.5& 2.5 \\
K2  & 7.4&  44.9& 0.2&  90.2& 2.5 \\
L1  & 6.3&  32.7& 1.1&  78.9& 2.0 \\
L2  & 2.0&   5.6& 0.1&  12.0& 2.8 \\
M1  & 0.2&   3.9& 0.0&   6.2& 3.0 \\
M2  & 14.0& 51.6& 0.2& 114.2& 2.4 \\
N   & 0.0&   0.0& 1.1&  34.4& 1.0 \\
O   & 4.3&  26.5& 0.1&  51.9& 2.5 \\

\enddata
\label{Corr}
\tablenotetext{a}{for definition of models see Table~\ref{models}}
\end{deluxetable}

\begin{deluxetable}{ccc}
\tablewidth{300pt}
\tablecaption{Predicted LIGO I Detection Rates (${\rm yr}^{-1}$)}
\tablehead{ Binary & Standard\hspace*{1cm} & Range  \\
            Type   & Model\hspace*{.5cm}   & (all models)}
\startdata

NS-NS& 1 $\times$ 10$^{-2}$ &  2 $\times$ 10$^{-4}$ -- 7 $\times$ 
10$^{-1}$ \\
BH-NS& 2 $\times$ 10$^{-2}$ &  2 $\times$ 10$^{-3}$ -- 7 $\times$ 
10$^{-2}$ \\
BH-BH& 8 $\times$ 10$^{-1}$   &  0 -- 2  \\
Total& 8 $\times$ 10$^{-1}$   &   2 $\times$ 10$^{-3}$ -- 2 \\
\enddata
\label{ligo1}
\end{deluxetable}

\begin{deluxetable}{ccc}
\tablewidth{300pt}
\tablecaption{Predicted LIGO II Detection Rates (${\rm yr}^{-1}$)}
\tablehead{ Binary & Standard\hspace*{1cm} & Range  \\
            Type   & Model\hspace*{.5cm}   & (all models)}
\startdata

NS-NS& 6 $\times$ 10$^{1}$ &  1 -- 4 $\times$ 10$^{2}$ \\
BH-NS& 8 $\times$ 10$^{1}$ &  9 -- 4 $\times$ 10$^{2}$ \\
BH-BH& 2 $\times$ 10$^{3}$ &  0 -- 8 $\times$ 10$^{3}$ \\
Total& 3 $\times$ 10$^{3}$ &  1 $\times$ 10$^{1}$ -- 8 $\times$ 
10$^{3}$ \\
\enddata
\label{ligo2}
\end{deluxetable}

\end{document}